\documentclass[onecolumn,superscriptaddress,preprint,showpacs,amsmath,amssymb]{revtex4-1}
\pdfoutput=1
\usepackage{booktabs,graphicx,pstricks,verbatim,url}
\usepackage{graphicx}
\usepackage{dcolumn}
\usepackage{bm}
\usepackage{slashed}
\usepackage{graphicx,bm,pstricks}
\usepackage{cancel}

\newcommand{\be}{\begin{equation}}
\newcommand{\ee}{\end{equation}}
\newcommand{\bear}{\begin{eqnarray}}
\newcommand{\eear}{\end{eqnarray}}
\newcommand{\bea}{\begin{eqnarray}}
\newcommand{\eea}{\end{eqnarray}}
\newcommand{\ba}{\begin{array}}
\newcommand{\ea}{\end{array}}
\def\ltap{\ \raise.3ex\hbox{$<$\kern-.75em\lower1ex\hbox{$\sim$}}\ }
\def\gtap{\ \raise.3ex\hbox{$>$\kern-.75em\lower1ex\hbox{$\sim$}}\ }

\newcommand{\lsim}{ \mathop{}_{\textstyle \sim}^{\textstyle <} }
\newcommand{\vev}[1]{ \left\langle {#1} \right\rangle }

\newcommand{\gev}{{\rm GeV}}
\newcommand{\tev}{{\rm TeV}}

\def\eg{{\it e.g.}}

\begin{document}

\preprint{FERMILAB-PUB-11-198-T}

\title{Higgs friends and counterfeits at hadron colliders}
\author{Patrick J. Fox} 
\affiliation{Theoretical Physics Department, Fermi National Accelerator Laboratory, Batavia, Illinois, USA}

\author{David Tucker-Smith}
\affiliation{Center for Cosmology and Particle Physics, Department of Physics, New York 
University, New York, NY}
\affiliation{Department of Physics, Williams College, Williamstown, MA 01267}
\affiliation{School of Natural Sciences, Institute for Advanced Study, Einstein Drive, Princeton, NJ 08540}

\author{Neal Weiner}
\affiliation{Center for Cosmology and Particle Physics, Department of Physics, New York 
University, New York, NY}
\affiliation{School of Natural Sciences, Institute for Advanced Study, Einstein Drive, Princeton, NJ 08540}
\date{\today}

\begin{abstract}
We consider the possibility of ``Higgs counterfeits''  - scalars that can be produced with cross sections comparable to the SM Higgs, and which decay with identical relative observable branching ratios, but which are nonetheless not responsible for electroweak symmetry breaking. We also consider a related scenario involving ``Higgs friends,'' fields similarly produced through $gg$ fusion processes, which would be discovered through diboson channels  $WW, ZZ, \gamma\gamma$, or even $\gamma Z$,   potentially with  larger cross sections times branching ratios than for the Higgs. The discovery of either a Higgs friend or a Higgs counterfeit, rather than directly pointing towards the origin of the weak scale, would indicate the presence of new colored fields necessary for the sizable production cross section (and possibly new colorless but electroweakly charged states as well, in the case of the diboson decays of a Higgs friend).
These particles could easily be confused for an ordinary Higgs, perhaps with an additional generation to explain the different cross section, and we emphasize the importance of vector boson fusion as a channel to distinguish a Higgs counterfeit from a true Higgs.  Such fields would naturally be expected in scenarios with ``effective $Z'$s,'' where heavy states charged under the SM produce effective charges for SM fields under a  new gauge force. We discuss the prospects for discovery of Higgs counterfeits, Higgs friends, and associated charged fields at the LHC.
\end{abstract}

\maketitle

\section{Introduction}
The origin of electroweak symmetry breaking (EWSB) has remained a mystery since the Standard Model (SM) was first written down. The appeal of a fundamental scalar Higgs lies in its simplicity, but brings with it the baggage of the radiative stability of the weak scale. At the same time, a strongly coupled mechanism for EWSB seems at odds with precision data, which point towards a light ($\ltap 200\ \gev$) Higgs, at which scale it is hard to imagine strong dynamics appearing without significantly affecting any number of observables.

The search for the Higgs, then, is not just an attempt to complete the SM, 
but rather a quest to understand the origin of the weak scale. 
Supersymmetry, strong dynamics, extra dimensions - 
in each case Higgs discovery (or non-discovery) would have important implications for the viability of the framework 
and would be a key predictor of what should come next.  
In light of this, it is essential not only to discover the Higgs, but to be certain that what we have discovered is, indeed, the field responsible for EWSB.

Imagine that a new resonance is discovered in one of the conventional channels -- such as $ b \bar b$, $W^+W^-$ or $\gamma \gamma$.  Even if the rate  differs significantly  from what is expected for the SM Higgs the resonance would potentially be hailed as the Higgs, perhaps with an added generation to explain the different cross section.  But how can we be certain that this is, in fact, the field that  generates the $W$ and $Z$ masses? Suppose we find the same state in a second channel, with the appropriate relative branching ratio to be the Higgs -- {\em then} can we  be sure that we've found the Higgs?

One might think that it would be difficult to imagine a scenario in which these signals 
are due to a  scalar that has little to do with EWSB, a Higgs counterfeit.   However, as we shall explore, the ingredients are are quite simple: an SM-singlet scalar, coupling to massive colored states, and a small mixing with the true Higgs via the Higgs portal. The ingredients are so simple, in fact, that this setup might arise in a variety of models.

More general than a Higgs counterfeit would be a ``Higgs friend,'' a field produced through conventional gluon fusion processes, but with branching ratios that can be very different than for a Higgs. 
These states might have negligible mixing with the Higgs, and therefore could
lack sizable tree-level couplings to SM fields.  In that case they would dominantly decay through loops of the fields that produced them (into $gg$ states), or through loops of related, electroweakly charged states, into $WW$, $ZZ$, $\gamma\gamma$ and $\gamma Z$ final states.  A Higgs friend could pop up quickly in SM Higgs search channels, but should be distinguished from a SM Higgs by the dramatic differences in signals in other expected channels.

The layout of this paper is as follows: in section \ref{sec:counterfeit}, we will explore the models that naturally give rise to such a Higgs counterfeit.   In this scenario a small amount of Higgs mixing gives rise to decay modes with observable branching ratios completely identical to those of the SM Higgs. 
In the Higgs counterfeit scenario there are tensions, however, in generating a suitably large cross section (from loops of colored states), with having a large BR into observable  final states. Building off these ideas, we then proceed to describe  ``Higgs friends'' in section \ref{sec:higgsbff}, particles produced through $gg$ fusion, but with branching ratios very different from the SM Higgs. 
With decay modes largely dominated by diboson channels, these friends might be found easily in ordinary Higgs search channels, but rather than pointing to the origin of EWSB, their discovery would point to new colored and electroweakly charged states. In section \ref{sec:effectivezprime}, we discuss how Higgs friends naturally arise in effective $Z'$ models, as have recently been discussed in the context of a variety of anomalies that have arisen at the Tevatron. In section \ref{sec:coloredstates} we consider the bounds on, and prospects for, the discovery of the new states - especially the colored ones - at the LHC. Finally, in section \ref{sec:discussion}, we conclude.  

Related work on the phenomenology of scalars which may be confused with the SM Higgs boson can be found in Refs.~\cite{Goldberger:1999un,Giudice:2000av,Csaki:2000zn,
Goldberger:2007zk,DeRujula:2010ys,Low:2010jp,Davoudiasl:2010fb}.

\section{A higgs counterfeit}
\label{sec:counterfeit}
To understand the role that Higgs counterfeits could play in upcoming searches, we essentially need to answer two questions: over general parameter ranges, what is the size of the signal that could be produced at the LHC, and how does this compare to that of the SM Higgs? 

Consider a real scalar field $S$, a singlet under the Standard Model (SM) gauge group.  This field can couple to the SM Higgs doublet $H$ via the Lagrangian terms $(\mu S +S^2) H^\dagger H$, leading to mixing between $S$ and the neutral component of $H$ once electroweak symmetry is broken.  The  mass-eigenstate  scalars, which we call $\phi$ and $\tilde{h}$, are then linear combinations of $h$, which has the same couplings to matter and gauge fields as a SM Higgs, and  a sterile state $s$, 
\bea
\phi&=& \cos\theta \;s - \sin\theta \;h \\
\tilde{h}&=& \cos\theta \;h+ \sin\theta \;s.
\eea
For small mixing ($\theta \ll 1$) $\phi$ has highly suppressed couplings to ordinary matter and gauge fields, while $\tilde{h}$ couples like a SM Higgs.   

If $S$ interacts exclusively through its couplings to $H$, then $\phi$ decays only through its $h$ component.   In this case the total width of $\phi$ is sensitive to the value of $\theta$, but $\phi$'s branching ratios are not.  They  are identical to the branching ratios of a SM Higgs of the same mass.  In particular,  $\phi$ decays  like a SM Higgs even if $\theta$ is very small, despite the fact that in this case $\phi$ has very little to do with electroweak symmetry breaking.   
Of course, if $S$ interacts only with $H$,  $\phi$ is also only {\em produced} through its $h$ component, so that its production cross section at colliders is suppressed by a factor of $\sin^2 \theta$ relative to that of a SM Higgs of the same mass.  

In this paper we consider the possibility that $S$ couples not just to $H$, but also  to additional colored fields $\Psi$ not contained in the SM. The $\phi$ production cross section can then be large even for small $\theta$, due to the contribution to gluon fusion from $\Psi$ loops. We take $\Psi$  to be charged only under color in this section, and consider the case where the  $\Psi$ fields are also charged under the electroweak gauge group in section \ref{sec:higgsbff}.  

The main  results of this section do not depend on the detailed properties of $\Psi$, but  below we consider benchmark scenarios in which $\Psi$ is a color-octet Majorana fermion.  We adopt $m_\Psi=350$ GeV and $m_\Psi = 500$ GeV as our benchmarks, and discuss the existing bounds on these scenarios in Section~\ref{sec:coloredstates}.

How does the coupling to extra colored states affect the decays of $\phi$?   The partial width of $\phi$ into gluons can change dramatically because the  $\phi \rightarrow gg$ amplitude gains a new contribution from the diagram with $\Psi$ in the loop, which competes with the contribution with the top quark in the loop.  Provided that $\Psi$ is charged only under color,  however, the partial widths of $\phi$ into all other final states of interest are  unaffected by the coupling to $\Psi$ at leading order.  For any final state $X$ besides $gg$, we therefore have
\begin{equation}
\Gamma_\phi (X) = \sin^2\theta \; \Gamma_h(X),
\label{eqn:width1}
\end{equation}
just as we'd have in the absence of $\Psi$~\footnote{This applies for $q {\overline q}$ final states provided  $\sin^2\theta$ is not so small that $\Psi$-induced two-loop contributions  to $\phi \rightarrow q{\overline q}$ become competitive with the  tree-level contribution.  For example, taking $\Psi$ to be a color-adjoint Majorana fermion with a weak-scale mass and order-one Yukawa coupling to $\phi$, we estimate that this translates roughly to $\sin^2 \theta > 10^{-5}$. Even if eqn.~(\ref{eqn:width1}) is not satisfied for $X=q {\overline q}$ final states, our final result, eqn.~(\ref{eqn:counterfeitmain}), is still a good approximation for other final states ($\gamma \gamma$, $ZZ$, $WW$), given that  the $\phi \rightarrow gg$ amplitude is much larger than the  two-loop contributions to $\phi \rightarrow q {\overline q}$ amplitudes.}.  Here and throughout, a $\phi$ subscript indicates that the quantity refers to the production or decay of a $\phi$ particle of mass $m_\phi$, as calculated in the theory with $S$ and $\Psi$, while an $h$ subscript indicates that the quantity refers to the production or decay of a Higgs boson of the same mass $m_\phi$, {\em as calculated within the SM}.  

Eqn.~(\ref{eqn:width1}) tells us that the ratios of branching ratios for $\phi$ decays will be the same as for a SM Higgs of the same mass.  The only exception to this is the $gg$ final state, which is not likely to be accessible at hadron colliders.  If the $\phi$ production cross section is comparable to that of an ordinary Higgs, a possibility even for very small mixing  once $\Psi$ particles are included, the $\phi$ particle could  be confused with a Higgs. In fact, depending on the masses of $\phi$ and ${\tilde h}$, it is possible that a dominantly sterile $\phi$ would be detected before ${\tilde h}$, the mass eigenstate more closely connected to electroweak symmetry breaking.   We call a dominantly sterile scalar with a sizable gluon-fusion production cross section  and partial widths that obey eqn.~(\ref{eqn:width1}) a  Higgs counterfeit.   

To explore the properties of Higgs counterfeits in quantitative detail, it is useful to compare  the rate for producing a particular final state $X \neq gg$ through $\phi$ production and decay, with the rate for producing the same final state through production and decay of a SM Higgs of the same mass.  The relevant ratio is
\begin{equation}
R_{\phi} = \frac{\sigma_\phi \times  B_\phi(X)}{\sigma_h \times  B_h(X)},
\label{eqn:counterfeit}
\end{equation}
where $\sigma_\phi$ is the $\phi$ production cross section and $B_\phi(X)$ is the branching ratio for $\phi \rightarrow X$, and $\sigma_h$ and $B_h (X)$ are the analogous quantities for a SM Higgs of the same mass.  The value of $R_{\phi}$ depends on $\theta$, $m_\phi$, and the properties of the colored $\Psi$ particles, but it is the same for all final states $X$ excluding $gg$.  

Given that gluon fusion is the dominant production mechanism for both $\phi$ and a SM Higgs, we have
\begin{equation}
\frac{\sigma_\phi}{\sigma_h} = \frac{\Gamma_\phi(gg)}{\Gamma_h(gg)}.
\label{eqn:xsec1}
\end{equation}
Furthermore, the total width of $\phi$ can be written as
\begin{equation}
\Gamma_\phi(total) = \Gamma_\phi (gg) + \sin^2\theta\;\left[\Gamma_h(total)-\Gamma_h(gg) \right],
\label{eqn:width2}
\end{equation}
which simply expresses the fact that the partial widths of a Higgs counterfeit into any channel except $gg$ is given by eqn.~(\ref{eqn:width1}). Using eqns.~(\ref{eqn:width1},  \ref{eqn:xsec1},  \ref{eqn:width2}),  eqn.~(\ref{eqn:counterfeit}) can be recast as
\begin{equation}
R_{\phi} = \frac{\sin^2\theta}{B_h(gg)+\sin^2\theta \; \frac{\Gamma_h (gg)}{\Gamma_\phi(gg)}\;(1-B_h(gg))}.
\label{eqn:counterfeitmain}
\end{equation}
Unless $\phi$ is lighter than a few GeV, the factor $(1-B_h(gg))$  in the denominator is roughly unity.   This expression for $R_{\phi}$ makes it clear that  to have a signal comparable to or larger than would be expected for a SM Higgs of the same mass, two conditions must be met:  $\sin^2\theta$ must not be much smaller than $B_h(gg)$, and $\Gamma_\phi(gg)$ must not be much smaller than $\Gamma_h (gg)$.  It is also evident from this formula that  increasing the $\phi$ production cross section with larger $\Psi$-loop contributions can only go so far in increasing the signal.  As the production cross section increases so does $\Gamma_\phi (gg)$, which eventually drives down branching ratios into interesting channels.   This is why $R_{\phi}$ saturates  at $\sin^2\theta/B_h(gg)$ for large $\Gamma_\phi(gg)$.  

For example, for $m_\phi \sim 120$ GeV  we have $B_h(gg) \sim 0.1$, so that a relatively large mixing $\sin^2\theta > 0.1$ is required to have a $\gamma \gamma$ signal from $\phi$ production and decay comparable to what a 120 GeV Higgs would give in the SM.   On the other hand, for $ m_\phi \sim 200$ GeV we have  $B_h(gg) \sim10^{-3}$, and large $ZZ$ and $WW$ rates are possible for much smaller mixing.  

We can do  for ${\tilde h}$ just what we did for $\phi$ in obtaining Eqn.~\ref{eqn:counterfeitmain}.  That is, we can calculate  $R_{\tilde h}$, the rate for producing a particular final state $X \neq gg$ through ${\tilde h}$ production and decay,  divided 
by the rate for producing the same final state through production and decay of a SM Higgs of the same mass.
We have
\begin{equation}
R_{\tilde{h}} = \frac{\cos^2\theta}{B_h(gg)+\cos^2\theta \; \frac{\Gamma_h (gg)}{\Gamma_{\tilde h}(gg)}\;(1-B_h(gg))},
\label{eqn:higgslike}
\end{equation}
which approaches unity as $\theta$ goes to zero given that $\Gamma_{\tilde h}(gg)$ approaches $\Gamma_h (gg)$ in the same limit.  If $\tilde{h}$ is the mostly active state ($\cos^2\theta>1/2$), then we have $R_{\tilde h}\approx  \Gamma_{\tilde h}(gg)  / \Gamma_h (gg)$.  Here we are making the mild assumption that  $B_h(gg) \ll  \Gamma_{h}(gg)  / \Gamma_{\tilde h} (gg)$ is satisfied, as will be the case for the benchmark scenarios we consider below.

\begin{figure}[htbp]
\begin{center}
\includegraphics[width=0.35\textwidth]{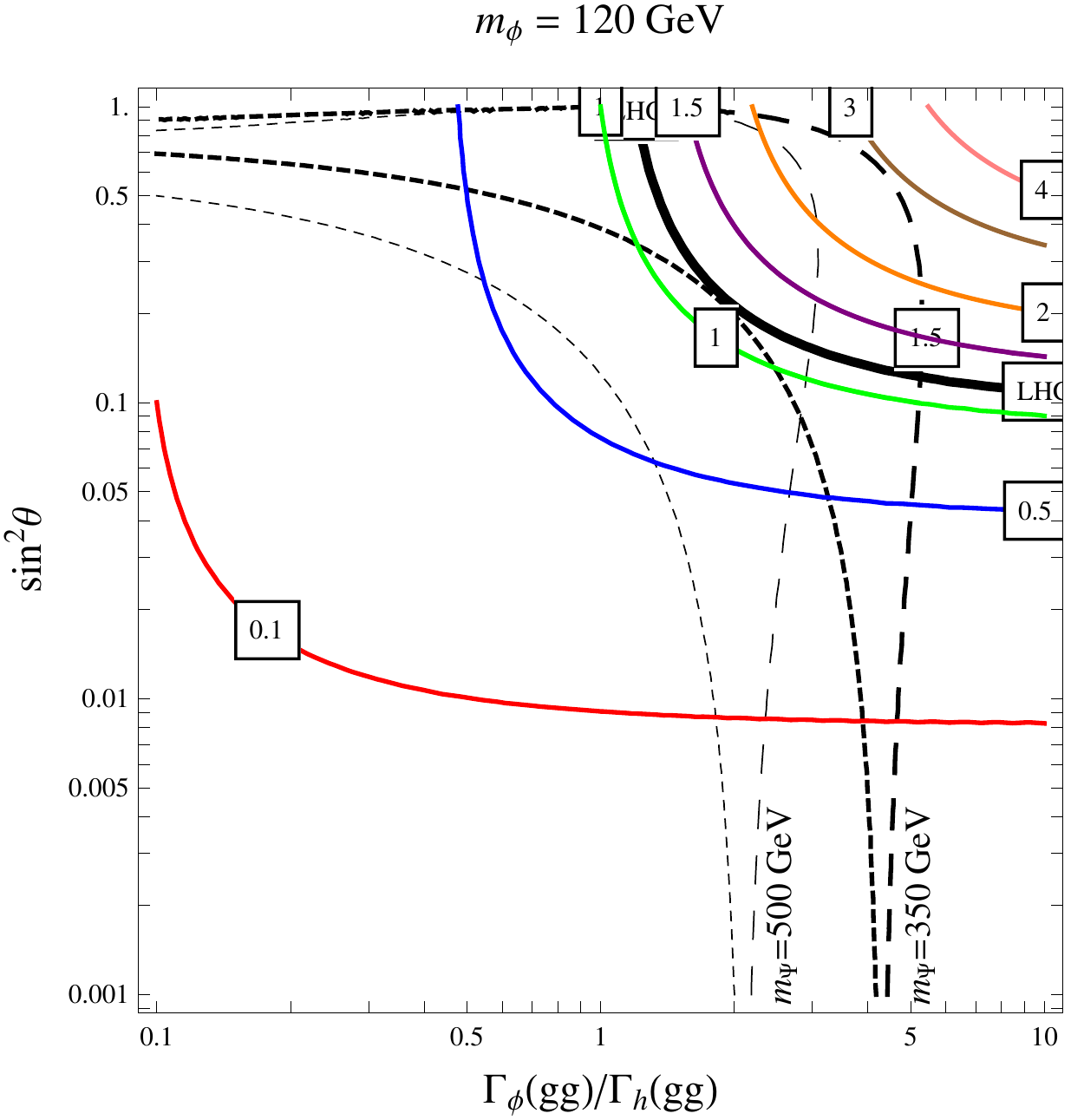}
\includegraphics[width=0.35\textwidth]{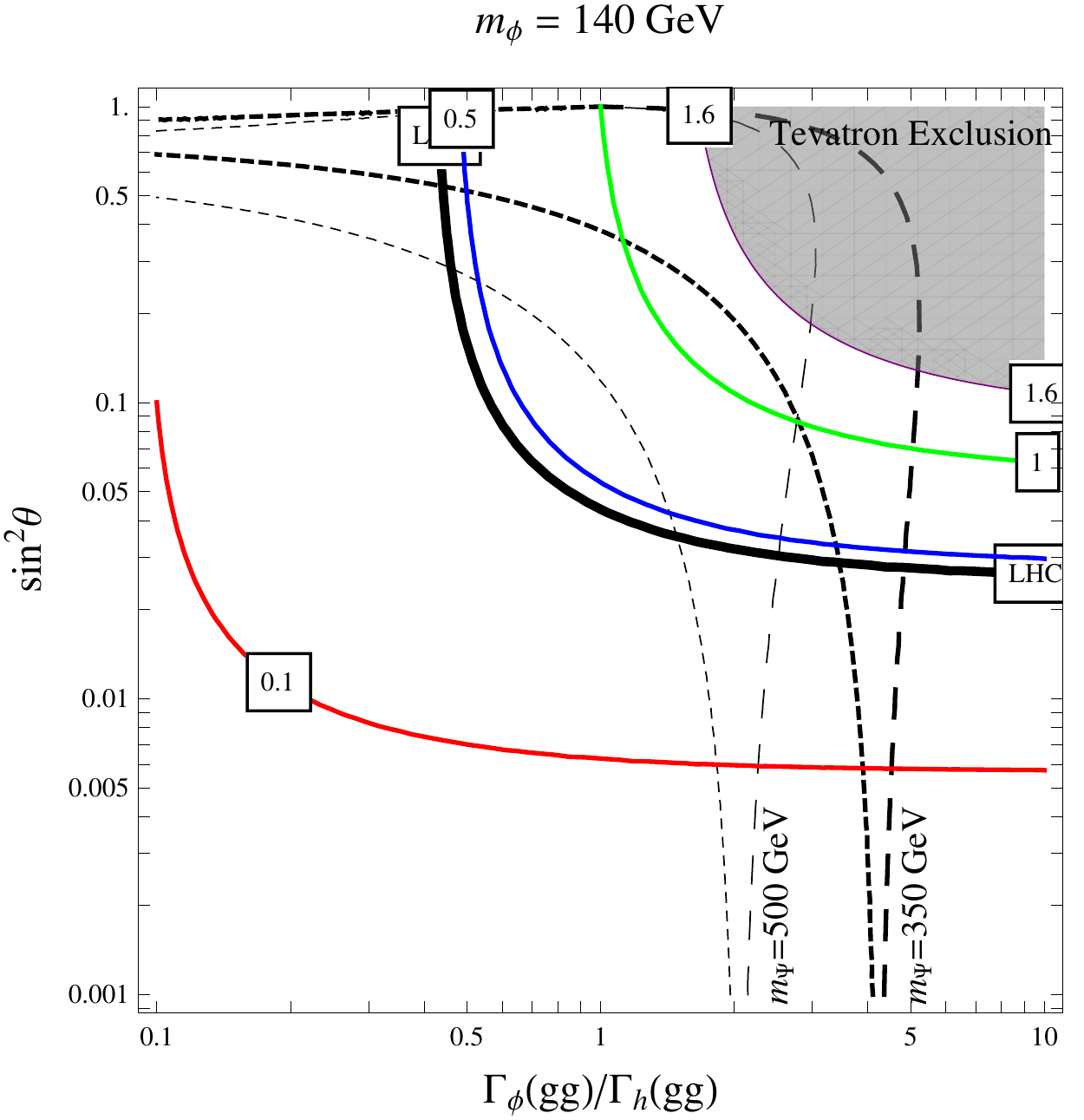}
\includegraphics[width=0.35\textwidth]{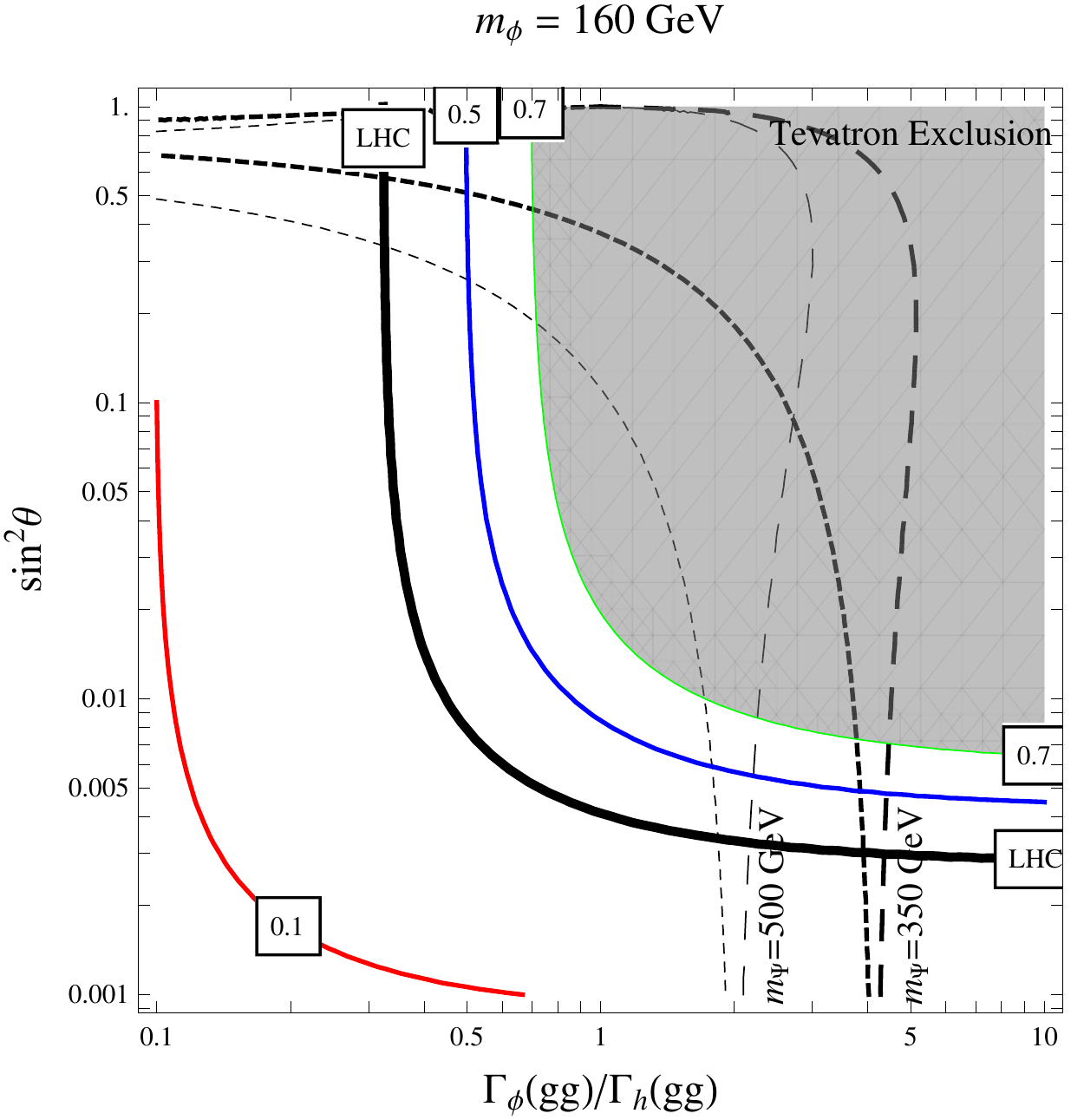}
\includegraphics[width=0.35\textwidth]{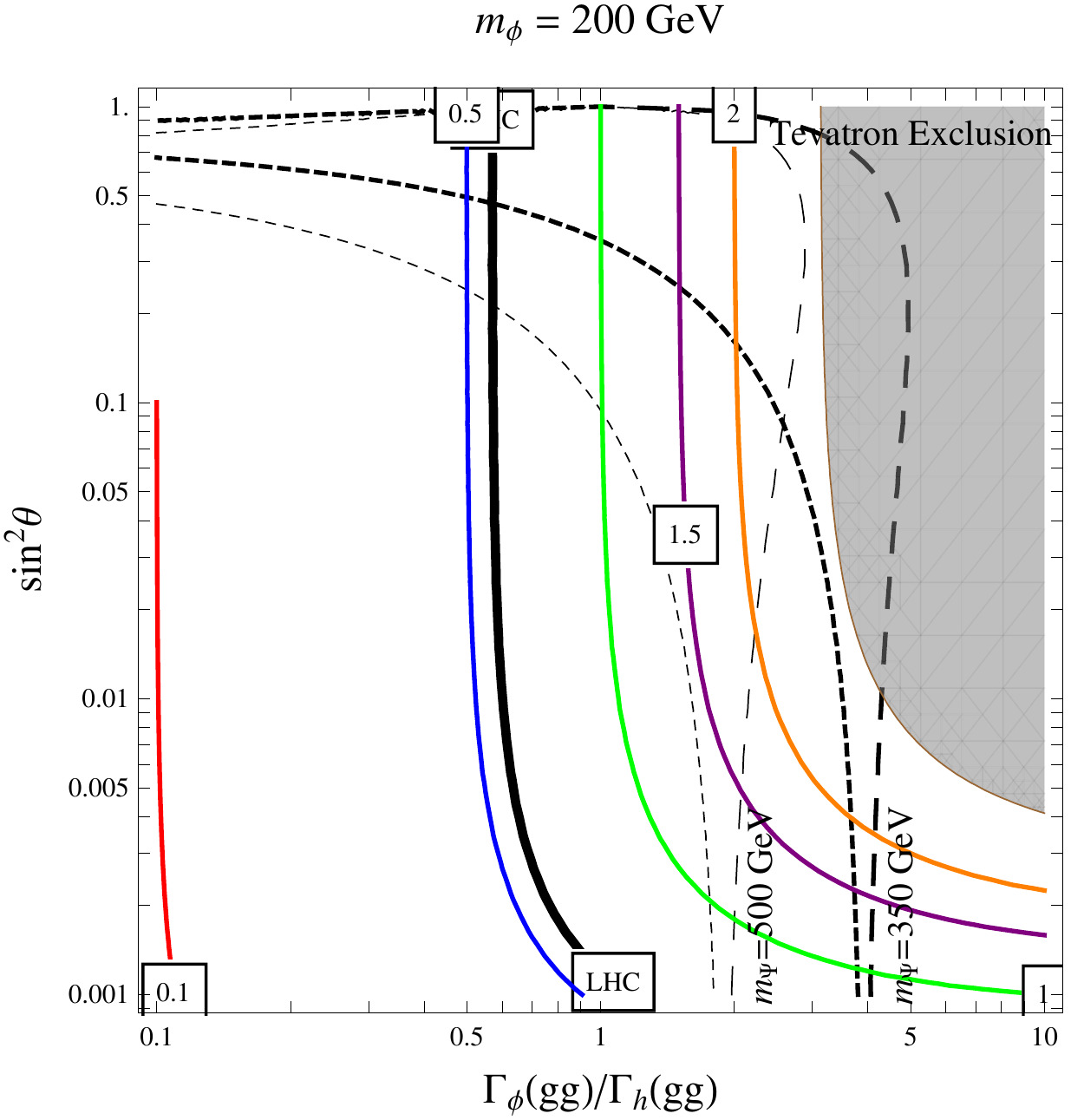}
\includegraphics[width=0.35\textwidth]{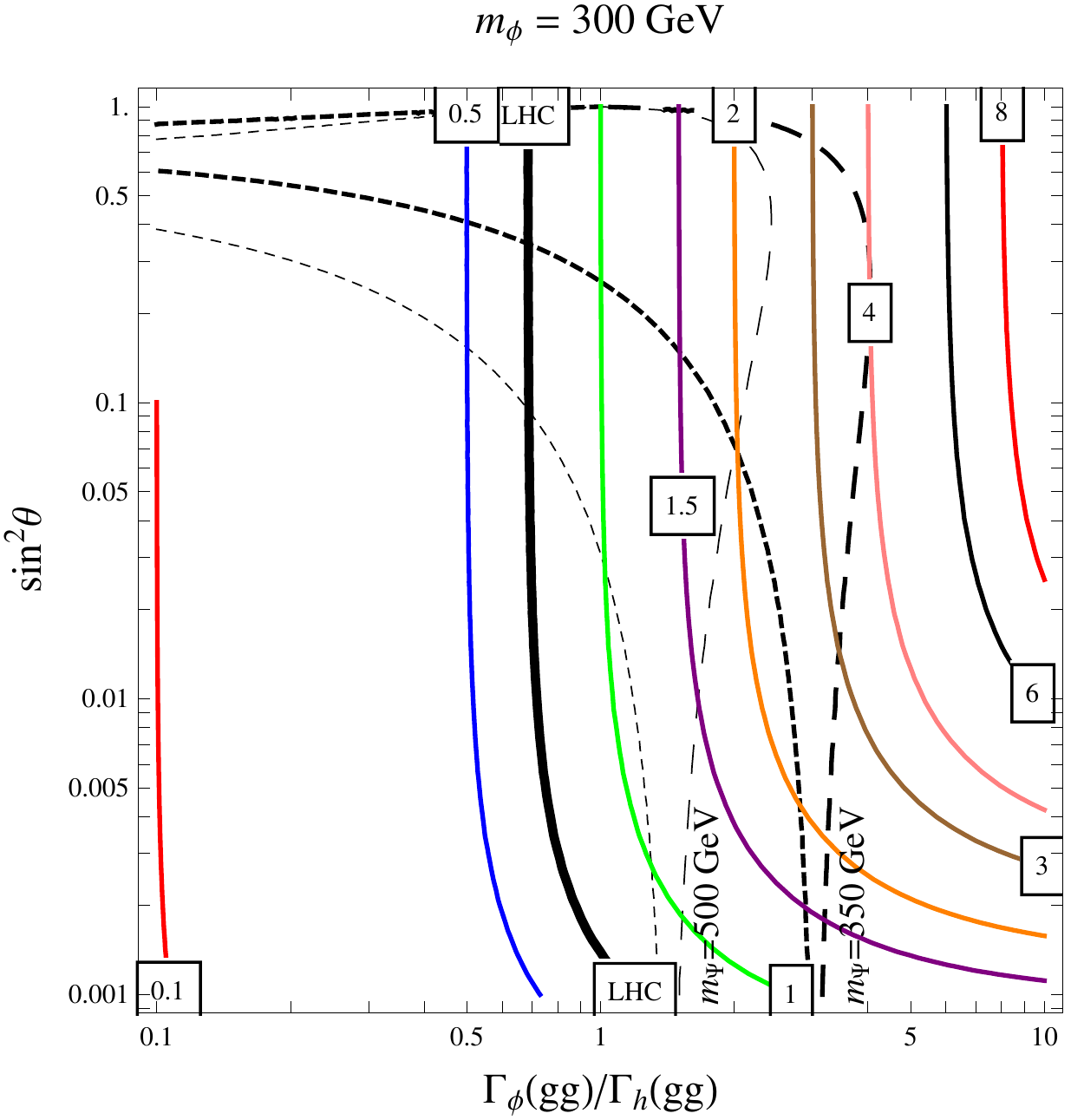}
\includegraphics[width=0.35\textwidth]{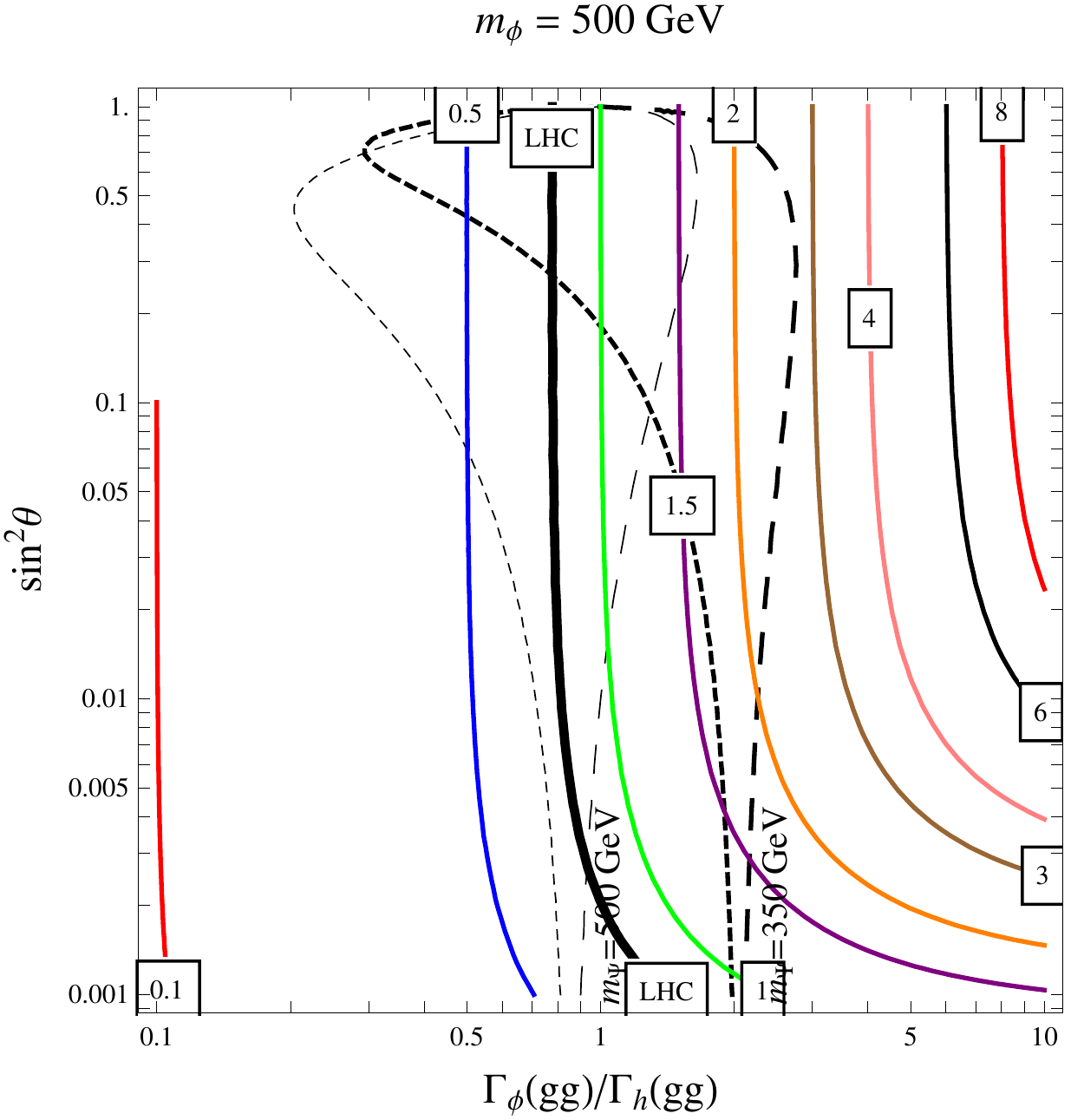}
\caption{Contours of $R_{\phi}$ (solid lines) for various $\phi$ masses.  Also shown is the region excluded by the Tevatron at 95\% CL (shaded), assuming  production only through gluon fusion, the projected 95\% CL  LHC sensitivity for 2 fb$^{-1}$ at 7 TeV  (thick, solid line),  and $\Gamma_\phi(gg)/\Gamma_h(gg)$ as a function of $\sin^2\theta$ for the benchmark case where $\Psi$ is a color-octet Majorana fermion with $y=1$, for $m_\Psi= 350$ GeV (darker) and $m_\Psi=500$ GeV (lighter).  The  benchmark contours have two branches: $\sin \theta<0$ (longer dashed) and $\sin \theta>0$ (shorter dashed). As $\sin^2\theta$ approaches 1, $\phi$ becomes very similar to a SM Higgs.}
\label{fig:fig1}
\end{center}
\end{figure}

\begin{figure}[htbp]
\begin{center}
\includegraphics[width=0.35\textwidth]{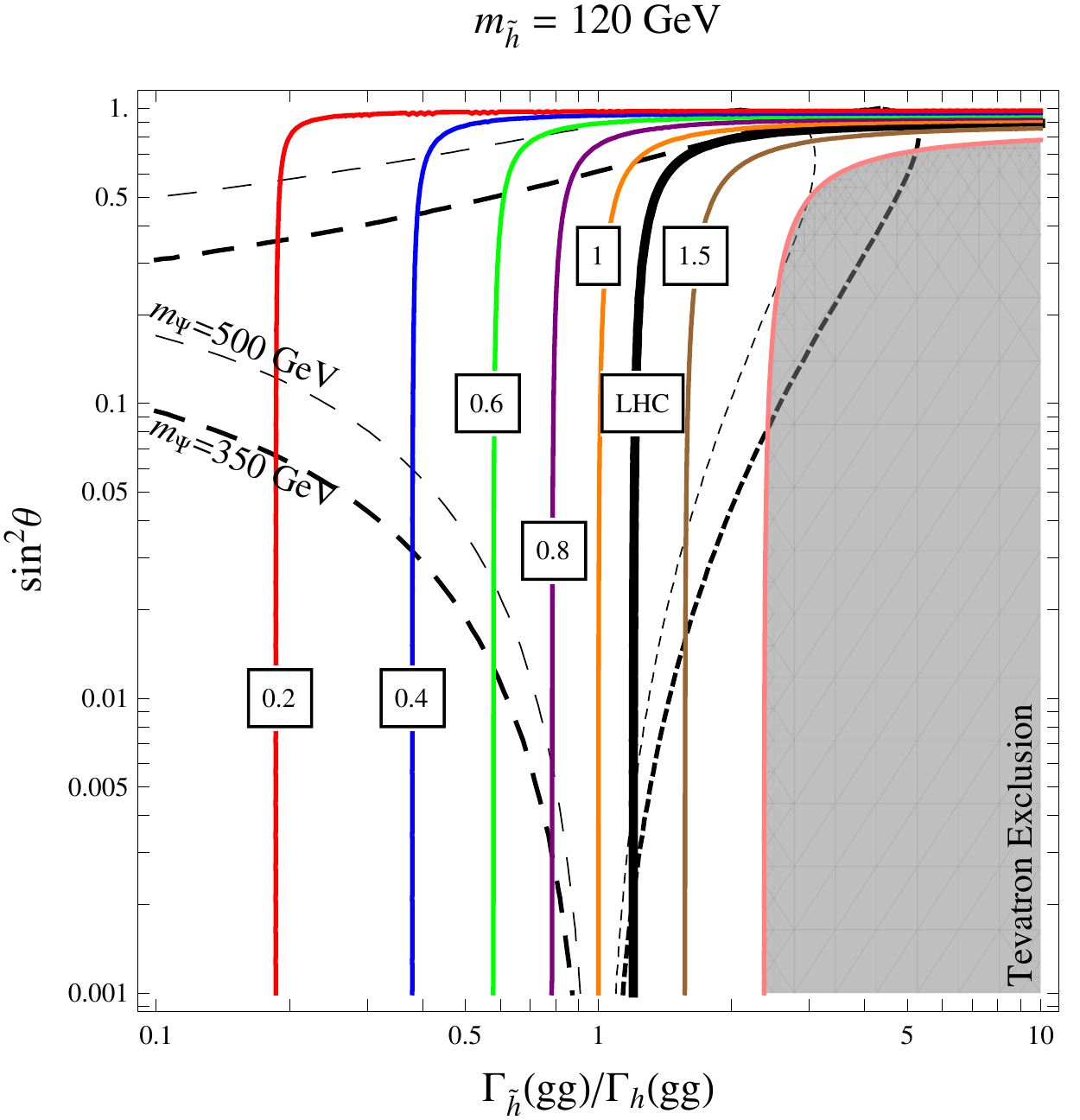}
\includegraphics[width=0.35\textwidth]{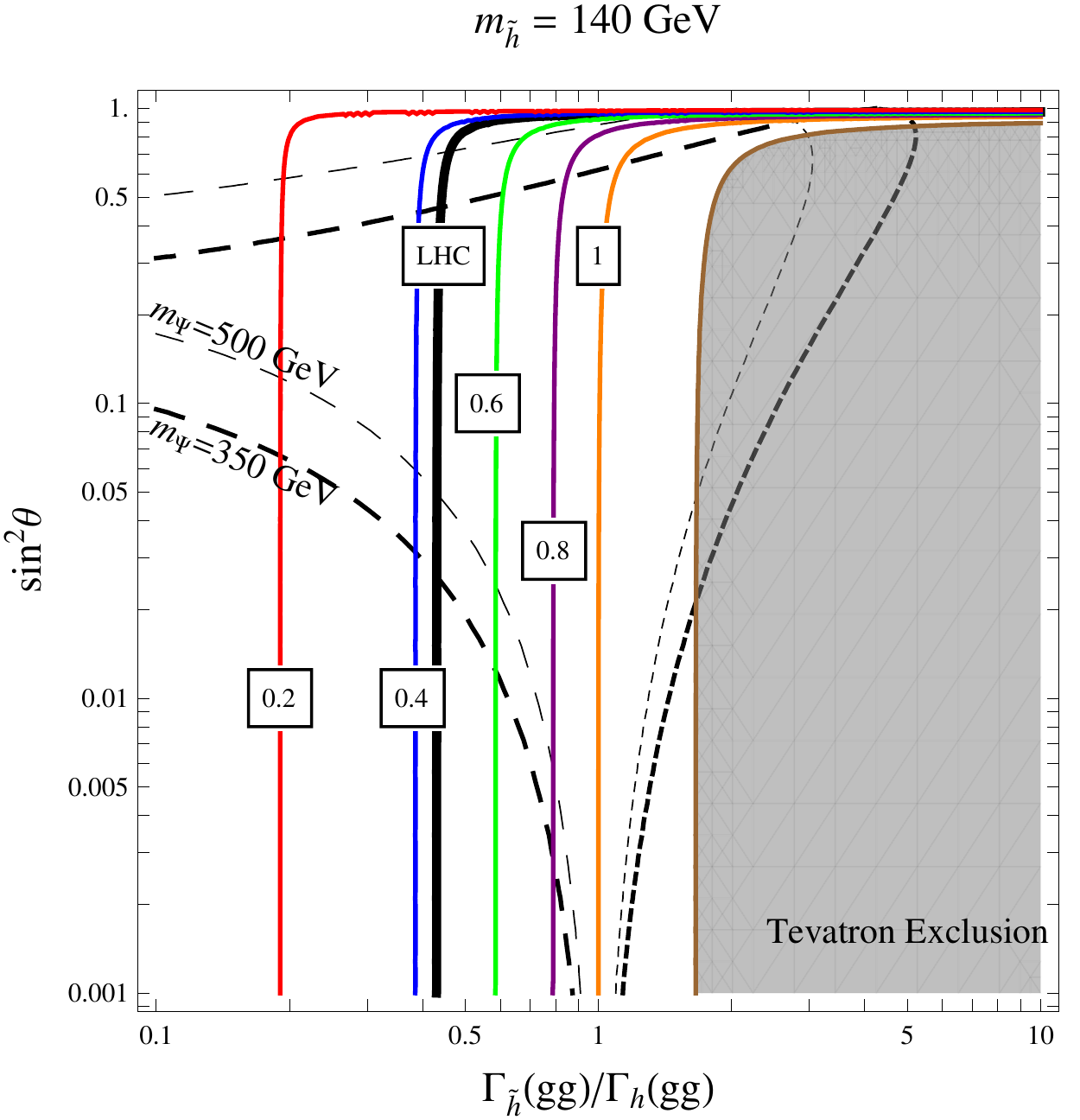}
\includegraphics[width=0.35\textwidth]{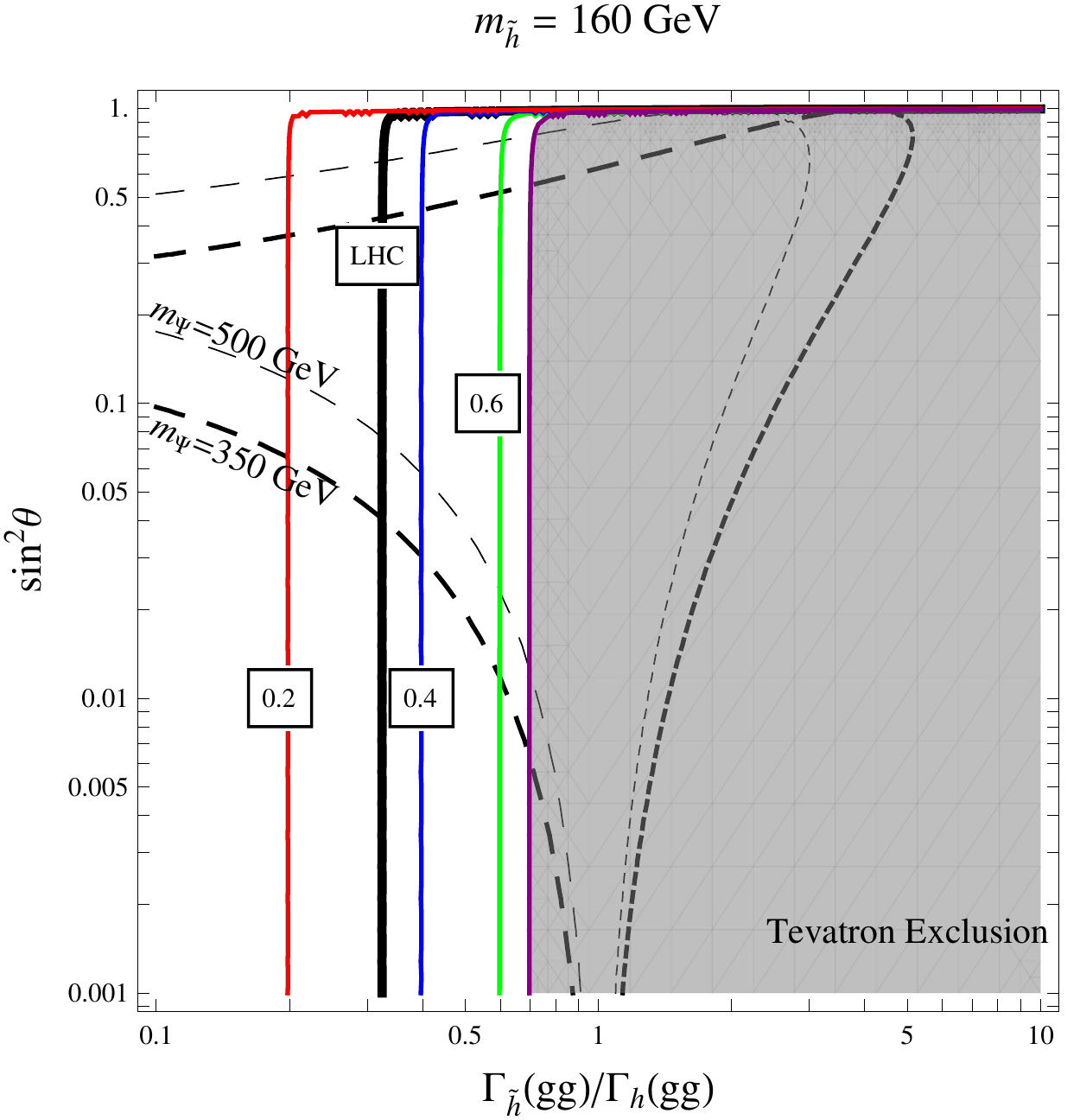}
\includegraphics[width=0.35\textwidth]{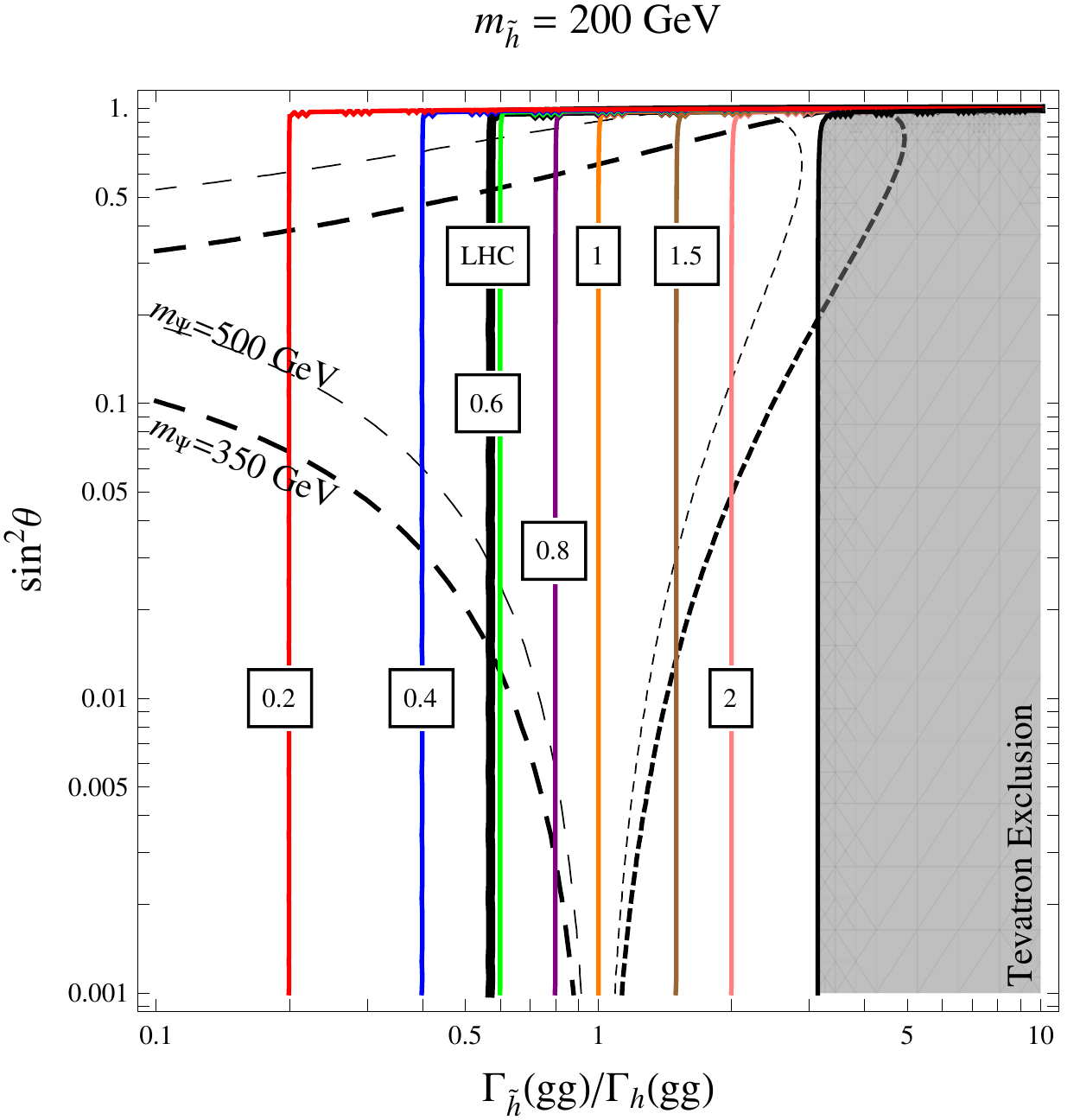}
\includegraphics[width=0.35\textwidth]{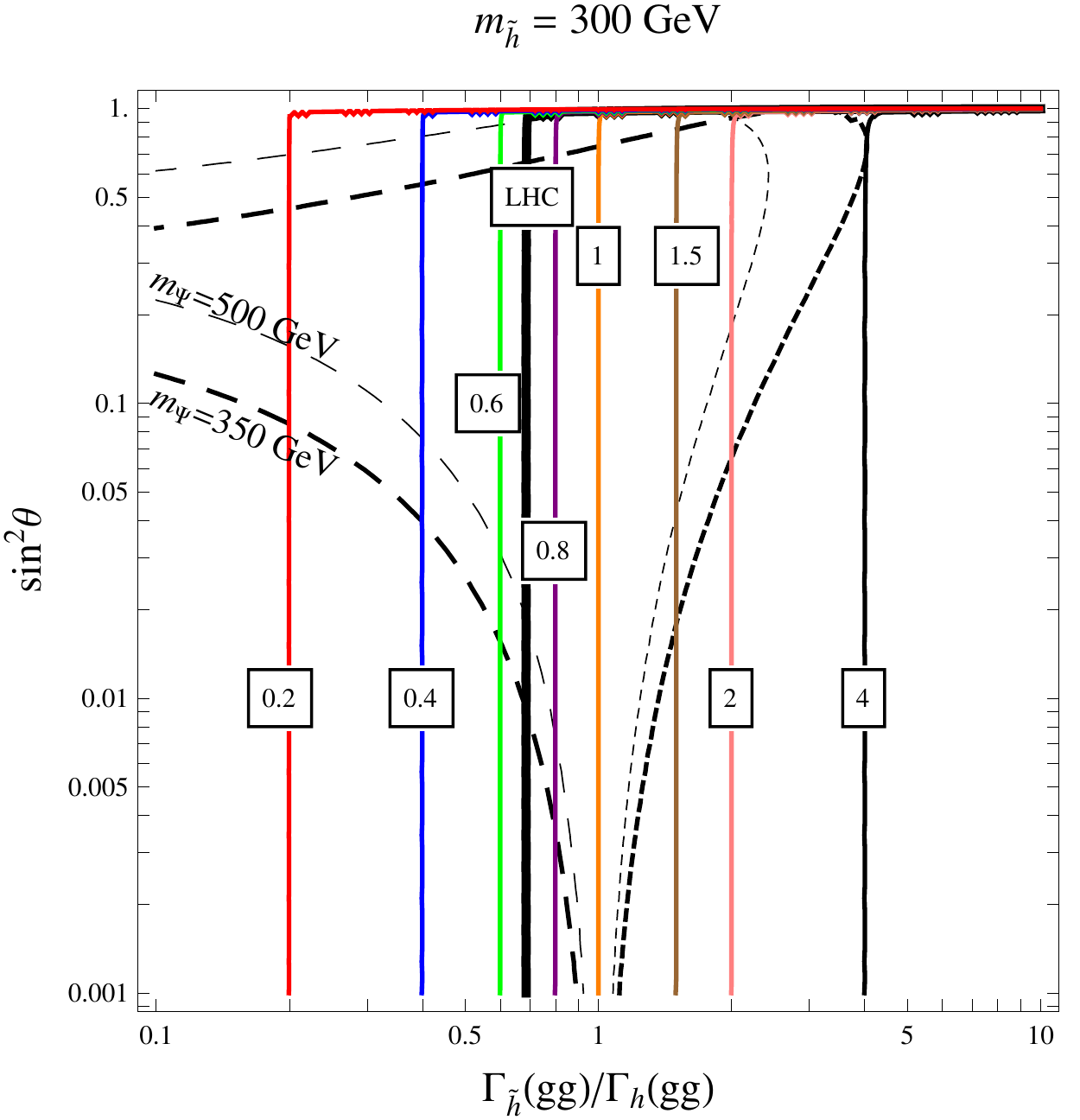}
\includegraphics[width=0.35\textwidth]{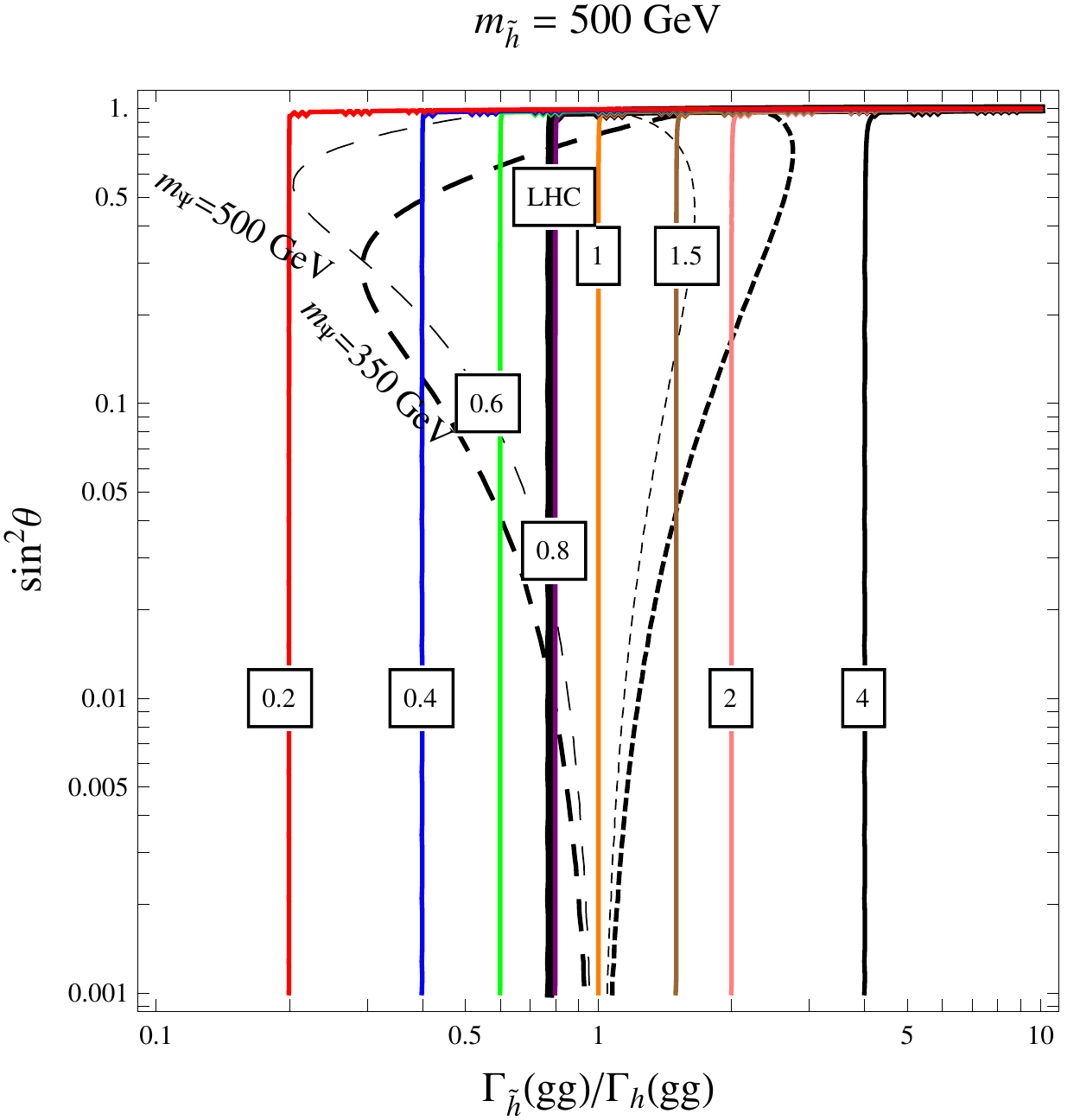}
\caption{Contours of $R_{\tilde h}$ (solid lines) for various ${\tilde h}$ masses.  The region excluded by the Tevatron at 95\% CL, using all production channels, is shaded.  Note that, for large $\sin^2\theta$ associated production turns off, but this correction is only an issue for $m_{\tilde{h}}=120$ GeV.  Also shown is the projected 95\% CL  LHC sensitivity for 2 fb$^{-1}$ at 7 TeV  (thick, solid line),  and $\Gamma_{\tilde h}(gg)/\Gamma_h(gg)$ as a function of $\sin^2\theta$ for the benchmark case where $\Psi$ is a color-octet Majorana fermion with $y=1$, for $m_\Psi= 350$ GeV (darker) and $m_\Psi=500$ GeV (lighter).  The  benchmark contours have two branches: $\sin \theta<0$ (longer dashed) and $\sin \theta>0$ (shorter dashed). As $\sin^2\theta$ approaches 1, ${\tilde h}$ becomes very similar to a SM singlet.   }
\label{fig:higgslike}
\end{center}
\end{figure}

We plot contours of $R_{\phi}$ and $R_{\tilde h}$ for various $\phi$ and $\tilde{h}$ masses in figures \ref{fig:fig1} and  \ref{fig:higgslike}.  
Because they are plotted in the $\Gamma_\phi (gg)/\Gamma_h(gg)-\sin^2\theta$ and  $\Gamma_{\tilde h}(gg)/\Gamma_h(gg)-\sin^2\theta$ planes, these contours are independent of the detailed properties of the extra colored states. To evaluate Eqns.~(\ref{eqn:counterfeitmain}) and (\ref{eqn:higgslike}) we use HDECAY \cite{Djouadi:1997yw} to calculate $B_h(gg)$ as a function of the scalar mass.  
In the same plots we also show  $\Gamma_\phi (gg)/\Gamma_h(gg)$ and $\Gamma_{\tilde h} (gg)/\Gamma_h(gg)$ for  benchmark scenarios where $\Psi$ is  a color-adjoint Majorana fermion whose coupling  to $s$ is equal to one~\footnote{
We normalize the coupling $y$ so that
$
\mathcal{L} \supset -\frac{y}{2} \; s {\Psi}^{\rm T} \mathcal{C} \Psi
$
when $\Psi$ is a  Majorana fermion, 
$
\mathcal{L} \supset -y \; s {\overline \Psi} \Psi
$
for a Dirac fermion,
$
\mathcal{L} \supset-\frac{y}{2} \; s \Psi^2
$
when $\Psi$ is a real scalar,  and
$
\mathcal{L} \supset -y \; s |\Psi|^2
$
for a complex scalar.  
We define $s$ so that $y$ is positive. 
},
 for $m_\Psi= 350$ GeV and $m_\Psi=500$ GeV.
We calculate these width ratios at leading order, including the top-loop and $\Psi$-loop contributions to the $\phi \rightarrow gg$  and $\tilde{h} \rightarrow gg$ amplitudes.  For a fermionic $\Psi$ particle in color representation $r$ we have 
\cite{Wilczek:1977zn,Ellis:1979jy,Rizzo:1980gz,Georgi:1977gs}
\begin{eqnarray}
\label{eqn:ggwidth}
\Gamma_\phi(gg)  & = &  \frac{\alpha_s^2}{256 \pi^3}m_\phi\left|-\sin\theta \left( \frac{m_\phi}{v}\right) A_{1/2}\left( \tau_{t \phi}\right)+ 2\sqrt{2} \; y \; n_{\Psi}\; C(r)\; \cos\theta  \left( \frac{m_\phi}{m_\Psi}\right)  A_{1/2}\left( \tau_{\Psi \phi}\right) \right|^2 \\
\Gamma_{\tilde h}(gg)  & = &  
\frac{\alpha_s^2}{256 \pi^3}m_{\tilde h}
\left|\cos\theta \left( \frac{m_{\tilde h}}{v}\right) A_{1/2}\left( \tau_{t {\tilde h}}\right)+ 2\sqrt{2} \; y \; n_{\Psi}\; C(r)\; \sin\theta  \left( \frac{m_{\tilde h}}{m_\Psi}\right)  A_{1/2}\left( \tau_{\Psi{\tilde h}}\right) \right|^2.
\label{eqn:higgslikeggwidth}
\end{eqnarray}
Here $v^2 = G_F^{-1}/(2\sqrt{2}) \simeq (174\; {\rm GeV})^2$, $\tau_{t\phi} =m_\phi^2/(4 m_t^2) $, $\tau_{\Psi\phi} =m_\phi^2/(4 m_\Psi^2)$, and similarly for  $\tau_{t {\tilde h}}$ and      $\tau_{\Psi {\tilde h}}$, $y$ is the coupling of $s$ to $\Psi$, $n_{\Psi}=1 \;(1/2)$ when $\Psi$ is  Dirac (Majorana),  and $C(r)$ is defined by ${\rm tr}(t_r^a t_r^b) = C(r) \delta^{ab}$ for color generators $t_r^a$, {\em e.g.} $C(8) = 3$ for a color adjoint and  $C(3)= 1/2$ for a color triplet. The function $A_{1/2}$, defined in Ref.~\cite{Djouadi:2005gi},  approaches $4/3$ in the heavy-loop-particle limit ($\tau\rightarrow 0$).  When $\Psi$ is a spin-zero particle we should take $n_{\Psi}=1 \;(1/2)$ for a complex (real) scalar, and in the second terms on the right-hand-side of both equtions  $A_{1/2}$ needs to be replaced by the function $A_0$, which is defined in Ref.~\cite{Djouadi:2005gj} and approaches $1/3$ in the heavy-$\Psi$ limit.  The generalization of Eqns.~(\ref{eqn:ggwidth}) and (\ref{eqn:higgslikeggwidth}) to the case where $\Psi$ consists of multiple particles with various masses, spins, couplings, and color representations is straightforward.  

For $\sin \theta < 0$ the top and $\Psi$ contributions interfere constructively for $\Gamma_\phi (gg)$ and destructively for $\Gamma_{\tilde h} (gg)$, and the opposite is true for $\sin \theta > 0$.  This is why the benchmark contours in Figs. \ref{fig:fig1} and \ref{fig:higgslike} have two possible values of $\Gamma_\phi (gg)/\Gamma_h(gg)$ or  $\Gamma_{\tilde h} (gg)/\Gamma_h(gg)$ at each value of $\sin^2 \theta$.
Without changing the masses $m_\phi$ and $m_{\tilde h}$, the sign of $\sin\theta$ can be flipped by reversing the sign of the $s h$ mass term in the Lagrangian (or equivalently, by leaving that term alone and  reversing the sign of $y$). For small values of $\sin^2\theta$ the top-loop contribution to $\phi \rightarrow gg$ and the $\Psi$-loop contribution to ${\tilde h} \rightarrow gg$ are suppressed, and the two values of $\Gamma_\phi (gg)/\Gamma_h(gg)$ and $\Gamma_{\tilde h} (gg)/\Gamma_h(gg)$ converge.

In Figs.~\ref{fig:fig1} and \ref{fig:higgslike} , we also show the parameter regions ruled out at 95\% CL by the combined Tevatron Higgs searches \cite{Aaltonen:2011gs}, along with contours projecting the  95\% CL sensitivity at the LHC after 2 fb$^{-1}$ at  7 TeV, taken from Ref.~\cite{atlas789tev}.
We do not take into account the possibility that $\phi$ and ${\tilde h}$ might give overlapping signals.  For example, if $m_\phi$ and $m_{\tilde h}$ both happened to be between $\sim140$ and 180 GeV, their combined $WW\rightarrow 2l 2\nu$ signal would be more significant than for either particle alone.

The plots in Fig.~\ref{fig:fig1} show that a 120 GeV Higgs counterfeit could be detected in a Higgs search in the $\gamma \gamma$ channel, for moderate mixing, while heavier counterfeits could give Higgs-like $ZZ$ or $WW$ signals even for $\sin^2\theta$ as small as $\sim 10^{-3}$. A dominantly sterile counterfeit might be discovered before the particle that approximately functions as the ``real'' Higgs.  To see this, consider the case where the counterfeit is heavy enough to have a large branching ratio into $ZZ^{(*)}$, while the dominantly active scalar is light, $m_{\tilde h} \sim 120$ GeV.    Recent ATLAS \cite{atlas789tev} and CMS \cite{cms789tev} studies estimate that after 2 fb$^{-1}$ at $\sqrt{s} =7$ TeV,  the median sensitivity for observing a 120 GeV SM Higgs will be  $\sim 1.6-2\,\sigma$. The projected sensitivity for heavier Higgs bosons is larger, \eg $\sim2.8\,\sigma-3.8\,\sigma$ for a 200 GeV SM Higgs.  Because the sensitivity at these higher masses is dominantly from gluon fusion followed by $h \rightarrow WW,\;ZZ$,  the sensitivity for a counterfeit would roughly be $R_{\phi}$ times that for a SM Higgs.  For $m_\phi = 200$ GeV, our benchmark contours show that $\phi$ can give SM-like rates, and thus $\sim3-4\, \sigma$ significance after 2 fb$^{-1}$, for mixing angles as low as $\sin^2\theta\sim 10^{-3}$.  Taking $\sin^2\theta=10^{-2}$, for instance, $R_{\phi}$ falls in the range $\sim 1.5 -3$ for our benchmarks, opening up the possibility of signals in excess of  $5\sigma$ after 2 fb$^{-1}$.  For larger masses, {\em e.g.} $m_\phi = 300$ or 500 GeV,  the projected LHC sensitivity for observing a SM Higgs is still $
\sim 2.5 -3\sigma$.  Our plots show that for these masses as well, even  relatively small mixing can yield highly significant signals after 2 fb$^{-1}$.  

In another interesting scenario $\phi$ could be mixture of a singlet and a very heavy Higgs.  The Higgs sector might even be strongly interacting, with $s$ coupling weakly to that sector to give a dominantly sterile, much lighter $\phi$ particle.  These possibilities are disfavored by precision electroweak measurements, but it is conceivable that extra states, perhaps including colored states,  cancel off the oblique corrections  from the Higgs sector.  Even if the width of the ``real'' Higgs is so large that the Higgs becomes unrecognizable as a particle,  $\phi$ could still be a narrow resonance.  Mistaking $\phi$ for a light elementary Higgs boson would in this case mean coming to a qualitatively incorrect understanding of electroweak symmetry breaking.

Fig.~\ref{fig:higgslike} shows that the mixing can suppress the observability of  $\tilde{h}$.  For example, with $m_{\tilde h}= 120$ GeV and $\sin \theta = -0.1$, our $m_{\Psi}=350$ GeV and $m_\Psi = 500$ GeV benchmarks give $36\%$ and $26\%$ reductions in the $\gamma \gamma$ signal relative to an  SM Higgs, respectively.  Much more dramatic suppressions are possible for larger mixing.   This effect makes it even more plausible that the dominantly sterile state might be seen first.  Of course, it is also interesting that the mixing  gives {\em enhanced} ${\tilde h}$ signals if the interference is constructive.

After discovery the counterfeit might  be mistaken for an excitation of the field principally responsible for electroweak breaking.  This would be a natural interpretation even if the rate were somewhat larger than  that expected in the SM.  The discrepancy  might be attributed, for example, to  extra quarks in the gluon-fusion loop, although in that case the $\gamma \gamma$ rate would not receive the same enhancement as for other final states.    If the extra quarks had both vector-like masses and Yukawa couplings to the Higgs, there would be sufficient freedom to account for the excess.

These examples highlight the importance of vector boson fusion (VBF) and associated production in establishing that a particle that decays like a Higgs boson really is a Higgs boson.    In the SM the $h W W$ and $h Z Z$ couplings lead to associated production of Higgs with gauge bosons and VBF signatures involving forward jets. For a Higgs counterfeit, the $\phi WW$ and $\phi ZZ$   couplings are suppressed by a factor of $\sin \theta$, and these signals  are likely to be strongly suppressed.

  For example, 
if   a 120 GeV $\gamma \gamma$ resonance were discovered, then for a SM Higgs one would also expect signals in $\tau \tau$ plus forward jets from VBF and  $b {\overline b}$ with leptons and/or missing energy from associated production. For Higgs counterfeits, VBF and associated production signals such as these are suppressed by a factor
\begin{equation}
\frac{\sin^2\theta}{1+ B_h(gg)\left(\frac{\Gamma_\phi (gg)}{\Gamma_h(gg)}\frac{1}
{\sin^2\theta}-1\right) },
\label{eqn:forwardjets}
\end{equation}
leading to much smaller and possibly unobservable signals. So, detection of the $\tau \tau$ and $b {\overline b}$ signals  with full strength would confirm  the discovery of a Higgs as opposed to a Higgs counterfeit.
For $\sim 2$ fb$^{-1}$ at  $\sqrt{s} = 7$ TeV center-of-mass energy, the sensitivities projected for $\tau \tau$ and $b {\overline b}$ in refs. \cite{atlas789tev,cms789tev} are  well below that projected for  $\gamma \gamma$ and $WW^{*}$.  So,  if a  $\gamma \gamma$ signal consistent with a light Higgs is observed in the coming year, we may have to wait until the LHC energy upgrade to rule out the Higgs counterfeit possibility.  
At 14 TeV center-of-mass energy and higher luminosity, ATLAS analyses suggest that  the $\tau \tau$  \cite{Aad:2009wy}  and  $b \overline{b}$   \cite{atlasassocprod} signals  may eventually have comparable significance to that from $\gamma \gamma$.

  Depending on the mass, if a  signal consistent with a heavier Higgs decaying through $Z Z^{(*)}$  were seen,  an observable  $WW^{(*)}$ signal might be expected in the SM, and vice versa. A Higgs counterfeit would in fact give the same ratio of $WW^{(*)}$ and $ZZ^{(*)}$ rates.  However, for a SM Higgs these signals would be accompanied by $WW^{(*)}$ and $ZZ^{(*)}$plus forward jets whereas for a Higgs counterfeit the same factor from eqn.~(\ref{eqn:forwardjets}) suppresses the VBF signal.  The projected significance after 5 fb$^{-1}$ at $\sqrt{s}=8$ TeV in the VBF $H\rightarrow WW^{(*)} \rightarrow 2l 2\nu$ channel, is smaller than the combined significance from other channels by about a factor of three or more over the entire mass range \cite{cms789tev}.  It is possible that including $H\rightarrow WW^{(*)} \rightarrow l \nu j j$ would lead to a more competitive VBF signal.  It is also possible that VBF analyses are presently not the top priority for the experimental collaborations, and that the sensitivity will improve once these analyses are given more attention.    In any case,  as for a $\gamma \gamma$ resonance, it will probably not be immediately clear after discovery whether a true Higgs or only a Higgs counterfeit has been found. 
  
Studies at 14 TeV and higher luminosity have found that VBF signals can potentially be as significant as those from gluon fusion. 
  In a CMS  analysis based on 30 fb$^{-1}$, VBF with $WW\rightarrow l \nu jj$ was found to be competitive with $ZZ$ above 300 GeV \cite{cms2003} and competitive with inclusive $WW^{(*)}$ for lower masses up until where $ZZ$ takes over at $\sim 190$ GeV \cite{cms2006}.   In an ATLAS study, VBF $H \rightarrow W W^{(*)}\rightarrow 2l2\nu$ was also found to be competitive with inclusive $WW^{(*)}$ up to $\sim 190$ GeV \cite{Aad:2009wy}.   It would be very interesting to know what  combined sensitivity can be achieved using all VBF processes, over the entire mass range including $\sim 200-300$ GeV.  We stress that signals with forward jets will be indispensable for establishing that the  particle discovered does in fact play a central role in electroweak symmetry breaking.

For counterfeits heavy enough to decay to on-shell $Z$ bosons, LHC measurements of the total width of $\phi$  could eventually be used to discriminate against a SM Higgs.  After 300 fb$^{-1}$ integrated luminosity, the resolution of the width would be better than 10\% for a SM Higgs with a mass greater than 250 GeV \cite{Gianotti:2000tz}.  So, it would ultimately be apparent in the counterfeit case that the resonance was much narrower than a SM Higgs.  The ratio of the total width of $\phi$ to the total width of a SM Higgs is
\begin{equation}
\frac{\Gamma_\phi}{\Gamma_h} = B_h(gg)\frac{\Gamma_\phi(gg)}{\Gamma_h(gg)} + \sin^2\theta (1-B_h(gg)).
\end{equation}
Given that  $B_h(gg)$ is less than $10^{-3}$ for masses above 200 GeV, this means that the width will be much smaller than in the SM for small mixing, and quite likely smaller than the experimental resolution. Of course, at such a large integrated luminosity we expect that it will already be clear from the suppression of signals with forward jets that the resonance is not a SM Higgs.

We conclude this section with a discussion of how  the benchmark contours of Fig.~\ref{fig:fig1} depend on the properties of $\Psi$.  Matters are complicated by the interference between the top- and $\Psi$-loop contributions to $\phi \rightarrow gg$, but in the limit of small $\sin^2\theta$ the top-loop contribution becomes negligible and the various scalings are easier to summarize.  

In the small $\sin^2\theta$ limit, $\Gamma_\phi (gg)/\Gamma_h(gg)$ is proportional to the square of the Yukawa coupling $y$ of $s$ to $\Psi$. For a 200 GeV $\phi$, lowering $y$ from 1 to $1/2$ (while keeping other parameters fixed)  thus shifts the intersection of the 350 GeV benchmark contours with the horizontal axis from $\Gamma_\phi (gg)/\Gamma_h(gg) \simeq 4$ to $\Gamma_\phi (gg)/\Gamma_h(gg)\simeq 1$.  Taking the mixing to be $\sin^2 \theta = 10^{-2}$, $R_{\phi}$ falls from $\simeq 3$ to $\simeq 1$.  

For $m_\Psi \gg m_{\phi}/2$, $\Gamma_\phi (gg)/\Gamma_h(gg)$ is approximately proportional to $1/m_{\Psi}^2$.  So, for the range of $\phi$ masses chosen for figure \ref{fig:fig1}, doubling the mass of the color adjoint from $m_\Psi = 350$ GeV to $m_\Psi = 700$ GeV has about the same effect as reducing the Yukawa coupling by half.  

Finally, Eqn.~(\ref{eqn:ggwidth}) shows that for a $\Psi$ particle in color representation $r$,  $\Gamma_\phi (gg)/\Gamma_h(gg)$ is  proportional to $n_\Psi^2 \;C(r)^2$.   For example, we get about the the same value of $\Gamma_\phi (gg)/\Gamma_h(gg)$ for a 350 GeV Majorana fermion color octet as for a 700 GeV Dirac octet with the same Yukawa coupling, or as for three 350 GeV Dirac-fermion quarks with  the same  Yukawas. However, if $\Psi$ includes particles that are  charged under the electroweak group, $\phi$ no longer decays as a Higgs counterfeit.  We consider this charged-$\Psi$ scenario next.\\


\section{The benefits of a Higgs friend}
\label{sec:higgsbff}
In addition to coloured states running in the loop the $\Psi$ fields may also carry electroweak quantum numbers, or there may be additional states that couple to $\phi$ which are only electroweakly charged.  As in the previous section $\phi$ will be produced through gluon fusion with the coloured states in $\Psi$ running in the loop, but whereas before the only way to make $\phi$ visible was to mix with the SM Higgs and acquire its decays, in this case there is the additional possibility that $\phi$ may now decay into SM gauge bosons through loops involving the electroweakly charged components of $\Psi$.  The branching ratios of these decays modes are not related to those of the SM Higgs, and this case we say that $\phi$ is a ``Higgs friend.''   It is possible that $\phi$ has virtually no mixing with the Higgs, or it  could be that $\phi$ both mixes with $h$ \emph{and} couples to new states which carry electroweak charge.

For a given $\phi$ decay channel, the difference between the counterfeit and friend cases depends on the relative sizes of the mixing-induced  and  loop-induced decay amplitudes.  In particular, the $\Psi$-loop contribution to $\phi\rightarrow \gamma\gamma$ can compete with the contribution from  $s-h$ mixing, since in the SM this decay is also loop induced.  On the other hand, for all but extremely small mixing the decays of $\phi\rightarrow ZZ/WW$ will not be substantially altered, by loop contributions, from what is expected from mixing alone.  Similarly the decays to SM fermions will be identical to the counterfeit case unless there is an abundance of electroweak charged states in $\Psi$.

A Higgs friend heavy enough to decay to WW or ZZ is unlikely to be confused with the SM Higgs of similar mass since the Higgs decay is tree level not loop.  However, both $\gamma\gamma$ and $\gamma Z$ final states are loop generated in the SM so a light Higgs friend (with or without mixing) has the potential to be confused with the SM Higgs.  For instance, for a light scalar $m_\phi\ltap 120\,\gev$ the visible decay would be to $\gamma\gamma$ and the cross section times branching ratio can easily be larger than that for the SM Higgs, leading again to the possible discovery of $\phi$ before the SM Higgs.  Or, more confusingly, $\phi$ may be discovered in $\gamma\gamma$ around the same time as a different mass Higgs is found in $\tau\tau$.

As a example of the Higgs friend scenario, we consider a benchmark of a pair of fermions transforming as a $10-\overline{10}$ of $SU(5)$ and coupling to $\phi$ with $y=1$.  For simplicity, and in order to be conservative about the photon branching ratio, we consider all colored components of the $10-\overline{10}$ to be degenerate in mass and the color-singlet to be decoupled, either by having a large mass or no coupling to $\phi$.  Since the $10$ contains $(q_L,\,u_R^c,\,e_R^c)$ of the SM, the production cross section of $\phi$ through gluon fusion will be as in the previous section.  However, unlike the Higgs counterfeit case the ratios of partial widths in visible channels will be altered from those of the SM Higgs.

We consider first a light friend, $m_\phi\sim 120\,\gev$.
One of the important discovery channels for a SM Higgs of the same mass is $\gamma\gamma$, and to see how the the $\gamma \gamma$ rate from production and decay of a Higgs friend compares with that due to a SM Higgs,  we introduce a  ratio  analogous to $R_{\phi}$ introduced earlier.
Here we make the simplifying assumption that the total width of $\phi$ is still well-approximated by Equation~(\ref{eqn:width2}).  This assumption is valid provided that $\Gamma_\phi(total)$ is either dominated by tree-level amplitudes induced by mixing, or by decays into gluons.  If a sufficient number of new  color-neutral states with electroweak charge are added, it is possible for the total width to instead be dominated by loop-induced decays into electroweak gauge bosons, for small enough mixing.  Assuming that this is not the case, we can use Equation~(\ref{eqn:width2}) to find
\be
R_{friend} =  \frac{\Gamma_\phi(X)/\Gamma_h(X)}{B_h(gg)+ \sin^2\theta \frac{\Gamma_h(gg)}{\Gamma_\phi(gg)} \left(1-B_h(gg) \right) }~,
\label{eq:Rfriend}
\ee
with $X = \gamma \gamma$ given that we are interested in the diphoton final state.  We see that when our approximation for $\Gamma_\phi(total)$ is valid, $R_{friend}$ is identical to $R_{\phi}$ but with $\sin^2\theta$ replaced by a ratio of widths.  Because $\phi \rightarrow \gamma \gamma$ is affected by the particles in the loop, and not just by the mixing, the numerator in $R_{friend}$ is not bounded by 1 as it was  is $R_{\phi}$.

\begin{figure}[t] 
   \centering
   \includegraphics[width=0.45\textwidth]{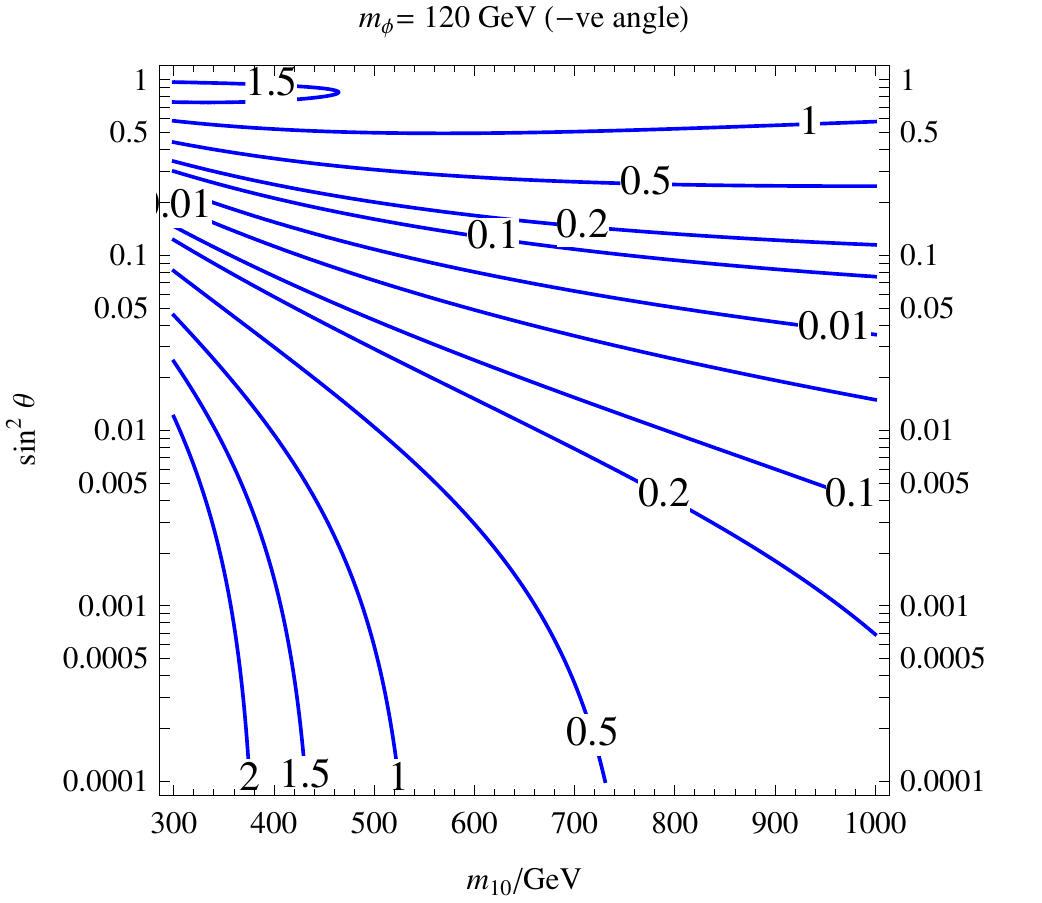} 
   \includegraphics[width=0.45\textwidth]{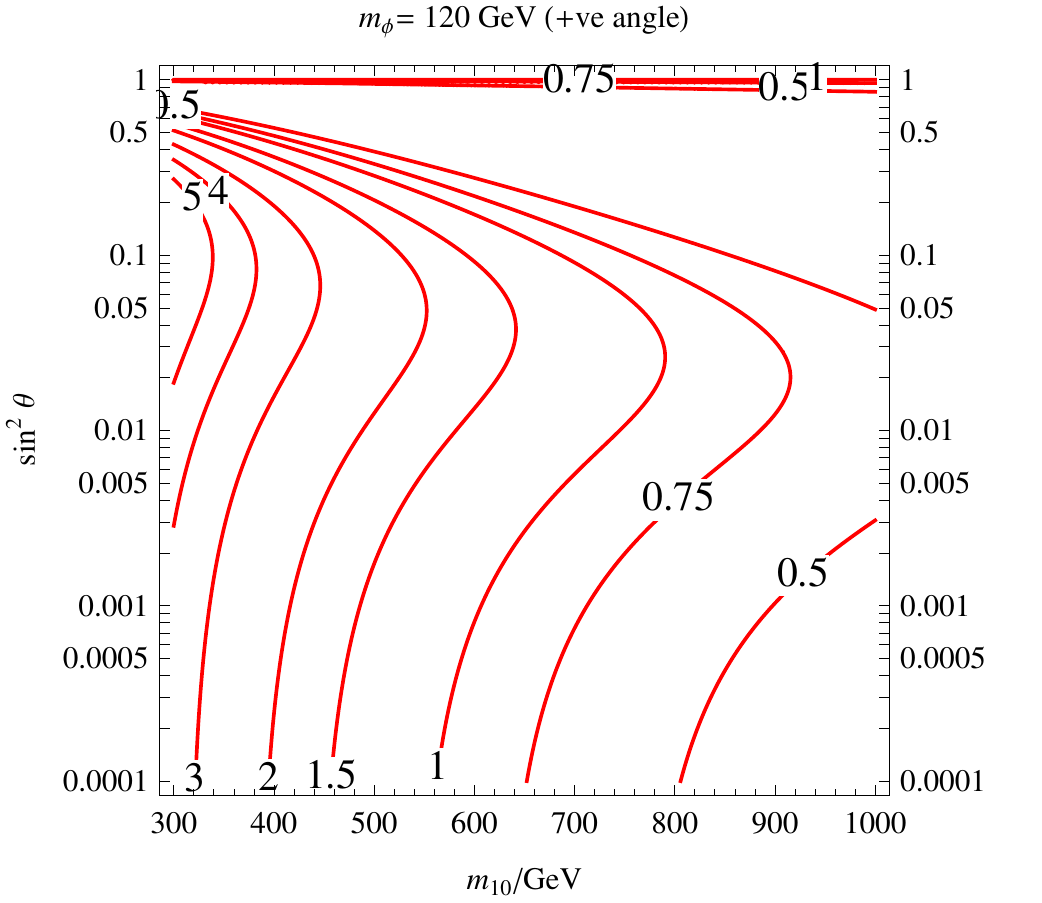}\\
   \includegraphics[width=0.45\textwidth]{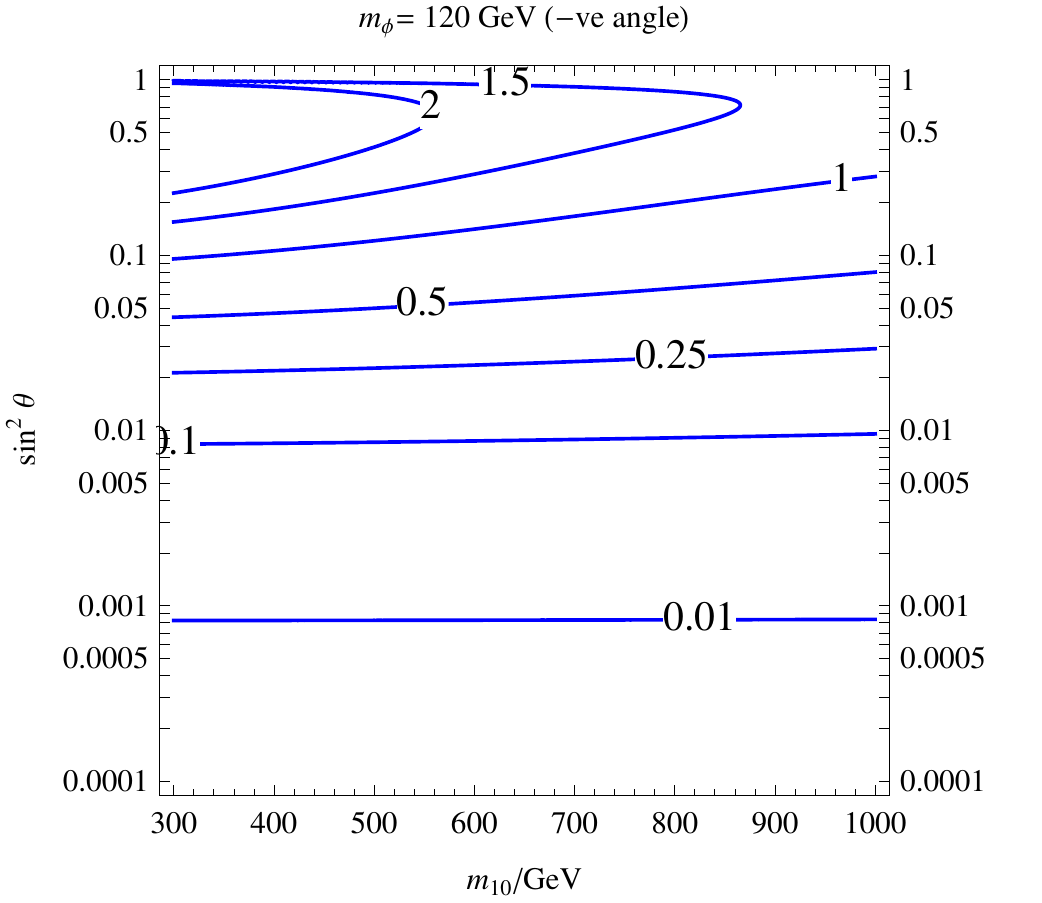} 
   \includegraphics[width=0.45\textwidth]{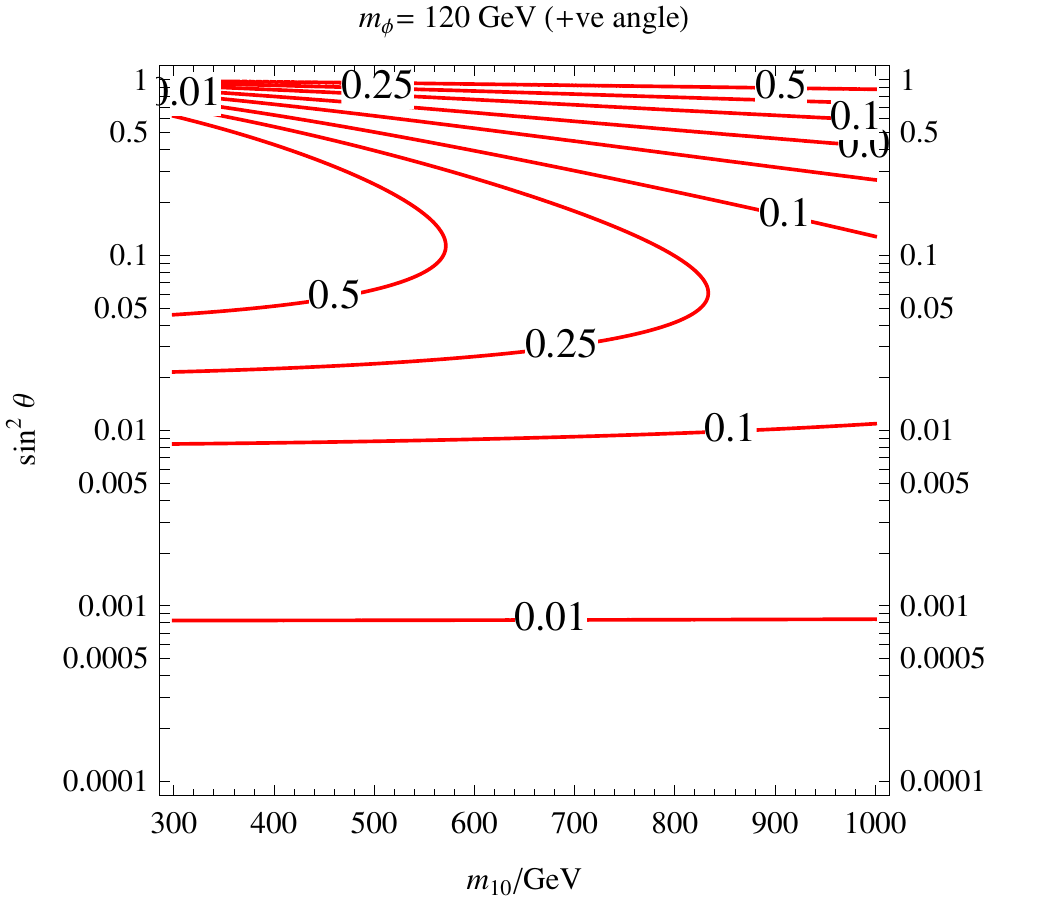}
   \caption{The ratio $R_{friend}$ for a $10-\overline{10}$ of mass $m_{10}$ (as discussed in the text, here we assume that only the colored components of the $10-\overline{10}$ pair couple to the friend).  The upper plots are for decays into $\gamma\gamma$ and the lower plots for $f\bar{f}$, in both cases the left-hand plots are for $\sin\theta<0$ and the right-hand plots are for $\sin\theta>0$.}
   \label{fig:Rfriends}
\end{figure}

For our benchmark case of a $10-\overline{10}$ of mass $m_{10}$ we show the behaviour of $R_{friend}$ in Figure~\ref{fig:Rfriends}.  As before we have used HDECAY to calculate the higher order corrections to the SM Higgs branching ratios, but use leading-order expression to calculate $\Gamma_\phi(\gamma \gamma)/\Gamma_h(\gamma \gamma)$ and  $\Gamma_\phi(gg)/\Gamma_h(gg)$.  Our expressions for  $\Gamma_\phi(gg)$  and $\Gamma_\phi(\gamma \gamma)$ appear in Equation~(\ref{eqn:ggwidth}) and in the Appendix, respectively.  As can be seen from (\ref{eq:Rfriend}) and Figure~\ref{fig:Rfriends}, even in the limit of no mixing $\phi$ has a potentially visible channel, $\phi\rightarrow \gamma\gamma$, whereas the $f\bar{f}$ final state becomes too small for $|\sin\theta|\ltap 0.1$.  Due to interference between SM and NP physics states running in the loop the behaviour depends on the sign of $\sin\theta$.
  
\begin{figure}[t] 
   \centering
   \includegraphics[width=0.7\textwidth]{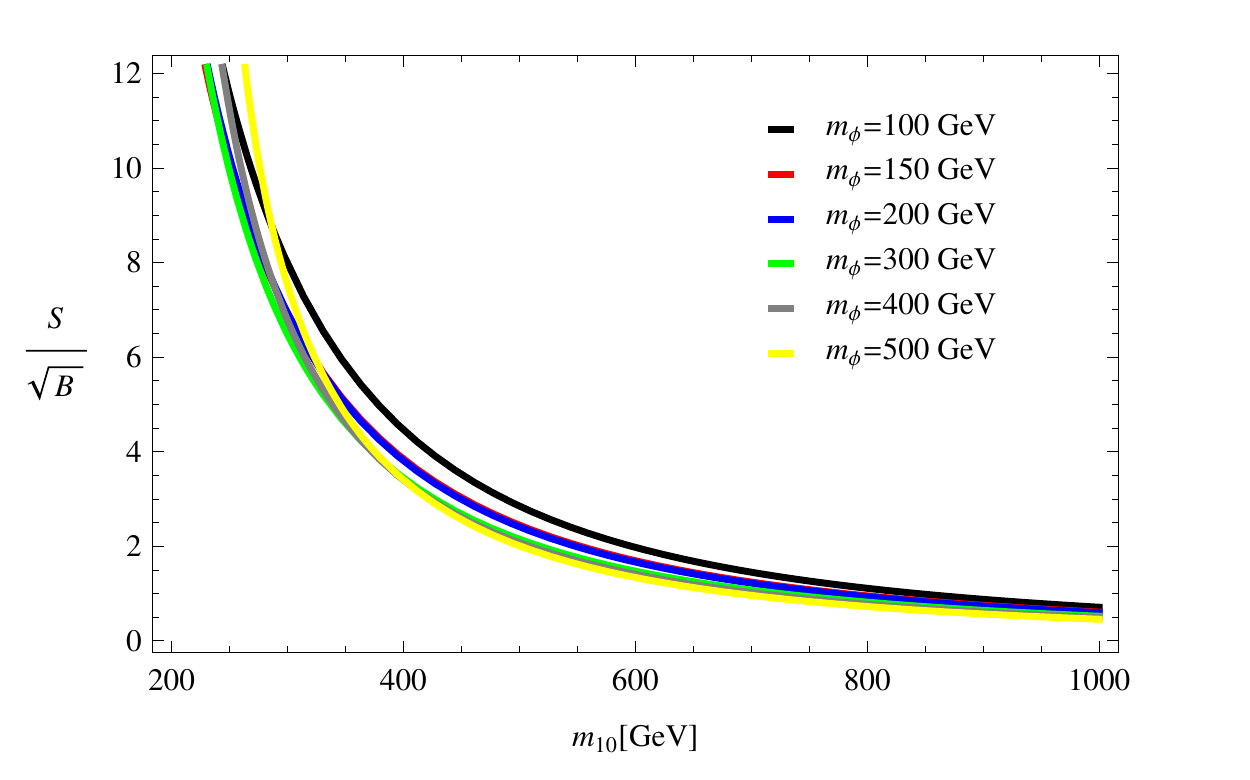} \\
   \includegraphics[width=0.45\textwidth]{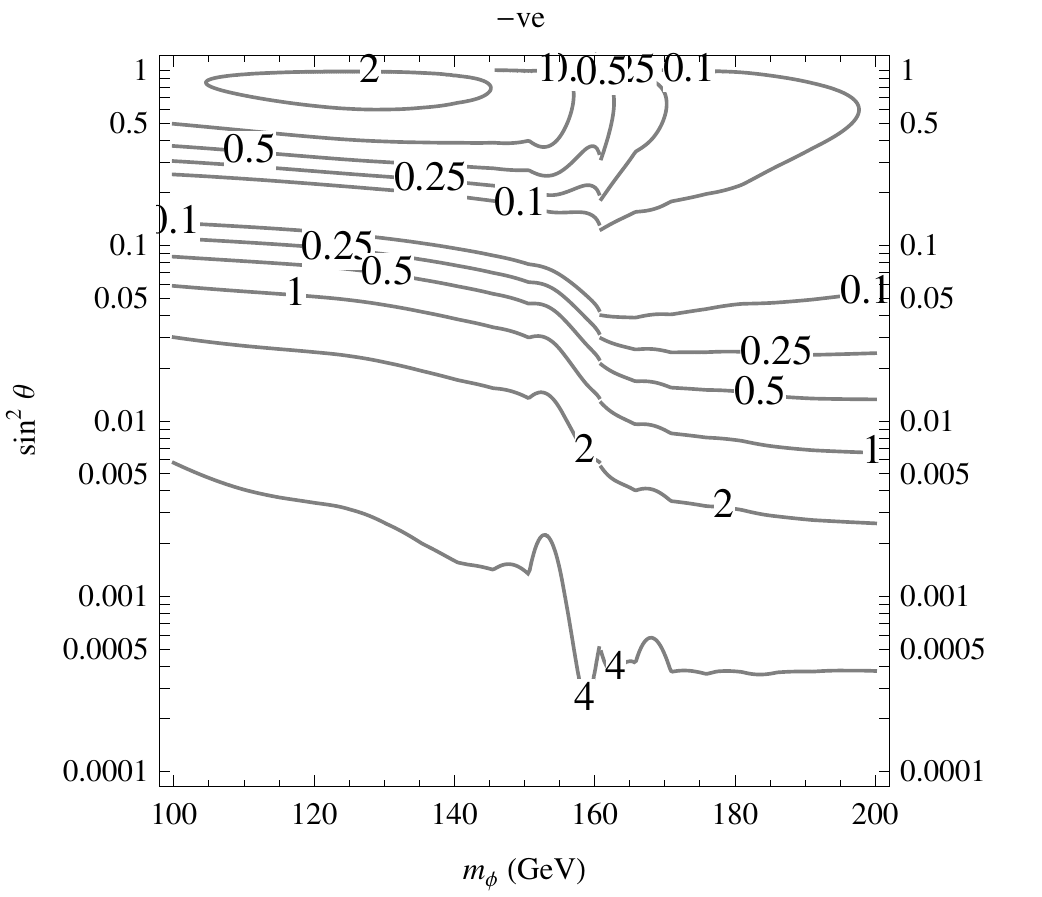} 
   \includegraphics[width=0.45\textwidth]{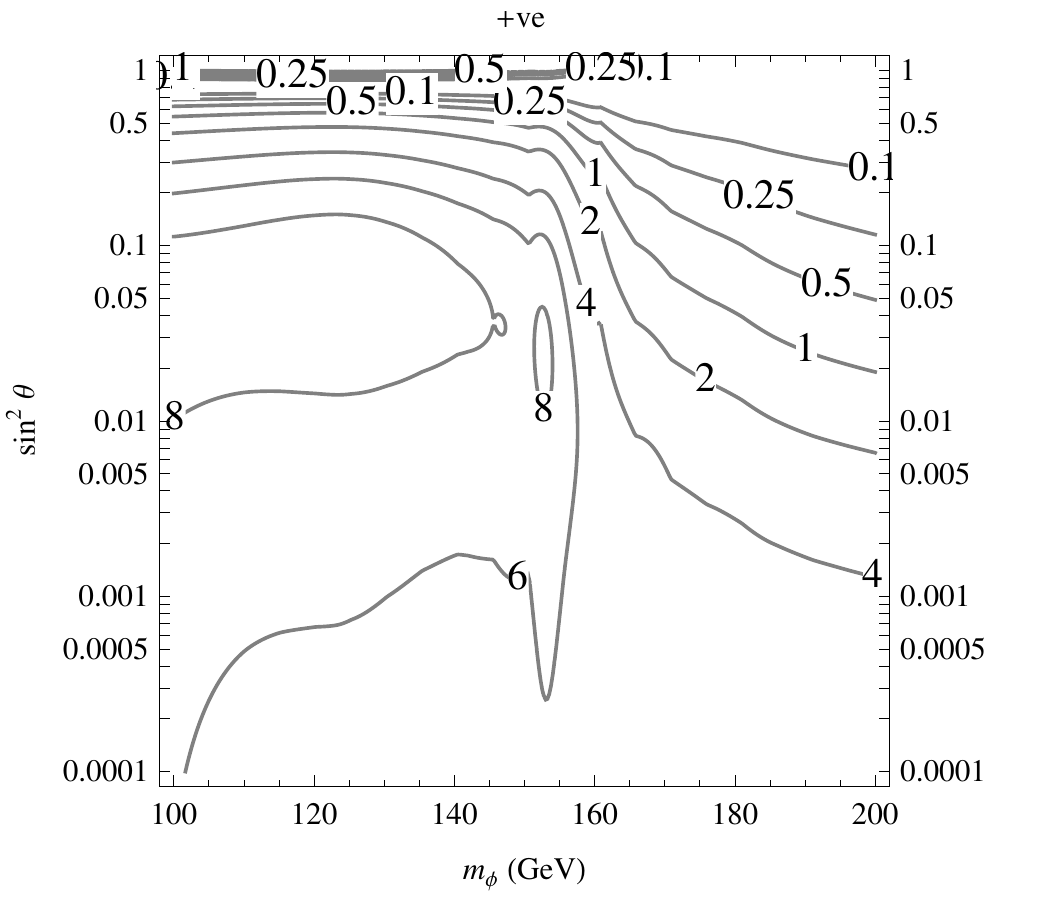} 
   \caption{The significance in $\phi\rightarrow\gamma\gamma$ after 10 fb$^{-1}$ at ATLAS, for $\sqrt{s}=14\ \tev$.  Top plot: Higgs friend with no mixing. Lower plot LH (RH): with mixing and $m_{10}=350\,\gev$ for negative (positive) mixing angle (as discussed in the text, here we assume that only the colored components of the $10-\overline{10}$ pair couple to the friend).}
   \label{fig:phisig}
\end{figure}

For a light SM Higgs the $\gamma\gamma$ channel has a branching ratio of $\sim 10^{-3}$, which is comparable to the branching ratio to photons in our benchmark scenario, if the Higgs friend does not mix, even for large $m_\phi$.  So it is possible that the Higgs friend may be found in the diphoton channel even at large masses.  To illustrate this point we show in Figure~\ref{fig:phisig} the signal significance for a Higgs friend produced through gluon fusion and decaying in the $\gamma\gamma$ channel, for our benchmark $\Psi$ of a $10-\overline{10}$.  We relate the significance for a Higgs friend of mass $m_\phi$ to a (light) SM Higgs of mass $m_h$ through,
\be
sig_\phi(m_\phi)=sig_h(m_h)\frac{(\sigma_\phi\times B_\phi)_{m_\phi}}{(\sigma_h\times B_h)_{m_h}}\sqrt{\frac{\sigma_{back}(m_h)\Delta E(m_h)}{\sigma_{back}(m_\phi)\Delta E(m_\phi)}}~,
\label{eq:significance}
\ee
where $\Delta E$ is the width of the diphoton peak, which we assume is proportional to the square root of the mass, and $\sigma_{back}$ is the cross section for background processes.
Using  $\sigma\times B(\gamma\gamma) \propto B(gg) B(\gamma\gamma)\Gamma(tot)$, 
we can rewrite the ratio of signal cross sections as
\be
\frac{(\sigma_\phi\times B_\phi)_{m_\phi}}{(\sigma_h \times B_h)_{m_h}} = \frac{(\sigma_h\times B_h)_{m_\phi}}{(\sigma_h\times B_h)_{m_h}}\left.\frac{\Gamma_\phi(gg)B_\phi}{\Gamma_h(gg)B_h}\right|_{m_\phi},
\label{eq:sigmabrrat}
\ee
where all branching ratios are for the $\gamma\gamma$ final state.  The first term in (\ref{eq:sigmabrrat}) is entirely a SM quantity and is calculated at NLO~\cite{Djouadi:2005gi} for  $s^{1/2}=14\,\tev$.  The expected significance, after 10 fb$^{-1}$ of 14 TeV data at ATLAS, of a light SM Higgs in the $\gamma\gamma$ channel is above 2 for the whole mass range in which this channel is usually considered, $110\, \gev \le m_h\le 140\,\gev$~\cite{Aad:2009wy}.   We consider only the irreducible background of $\gamma\gamma+X$.  This was calculated at NNLL~\cite{Balazs:2007hr} for di-photon invariant masses up to 250 GeV.  Their result is well fit by a quadratic in log-log space which allows us to extrapolate up to higher invariant masses.   In Figure~\ref{fig:phisig} we show the significance for a Higgs friend with and without mixing after 10 fb$^{-1}$ of 14 TeV data at ATLAS.

From  Figure~\ref{fig:Rfriends} we see that that one can achieve $R_{friend} \sim 5$ for our benchmark scenario, with $m_{10} \approx 350 \;\gev$ and $y=1$, and a natural question is then how large can be achieved more generally? Note that we have assumed the $e_R$ in the ${\bf 10}$ is decoupled from the $\phi$.  Including the $e_R$ can enhance $R_{friend}$ further, for example if the $e_R$ has a Yukawa coupling to $\phi$ equal to 1 and a mass of 100 GeV, then we find  $R_{friend}$ can be as large as 15, and is equal to $\simeq 13$ even in the absence of mixing.

We would also like to estimate the possible signals for other representations. Taking (\ref{eq:Rfriend}) in the $\sin^2\theta \rightarrow 0$ limit (i.e., a pure friend), 
we have
\be
R_{friend} \approx \frac{\Gamma_\phi(X)}{\Gamma_h(X) B_h(gg)}~.
\label{eq:Rfriendsimp}
\ee
Thus, we can compare the relative signals of different representations just by looking at $\Gamma_\phi(\gamma \gamma)$ (recall that the gluon width cancels in the production and branching ratios). For a $d^c$, the signal would be a whopping 81 times smaller than a comparably massed benchmark case. Completing it as a $\bf 5$ (with the addition of an $L$) produces a signal roughly 5 times smaller, as does a $u^c$ (which is part of the $\bf 10$), requiring much smaller masses to get a comparable signal for the same coupling. Of course, larger couplings would increase the signal.

Returning to the benchmark scenario with  only the heavy quarks coupling to $\phi$, from Figure~\ref{fig:phisig} it is clear that at small mixing, or masses below the $ZZ/WW$ threshold, there is potential for an early discovery in this channel.  For the no-mixing case $\phi$ can be discovered after 10 fb$^{-1}$ for $m_{10}\ltap 350\,\gev$ for all $m_\phi\le 500\,\gev$.  This independence of $m_\phi$ is because both the background and signal cross sections are falling as $m_\phi$ is increased and the branching ratio is remaining relatively constant.  This is to be contrasted with the mixing case where the existence of tree-level decays into massive vector bosons can compete with the loop induced decays, if $|\sin\theta|\gtap 0.01$, lowering the significance in this channel.  Although $\gamma\gamma$ may no longer be the discovery channel it is still an interesting channel to look in since measurement of $B(\gamma\gamma)$ will distinguish the friends case from the counterfeit case.
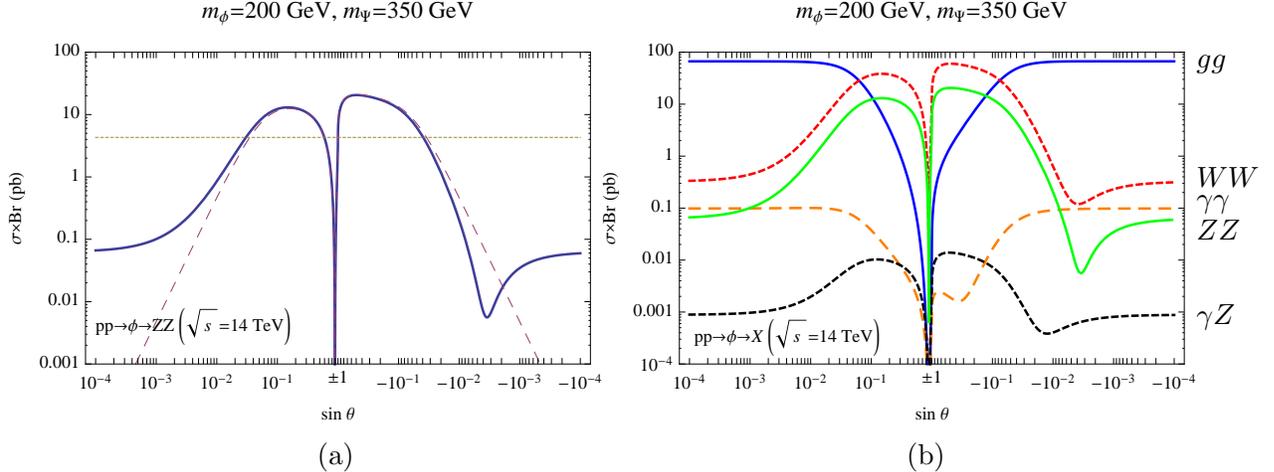
\begin{figure}[t] 

   \centering

   \includegraphics[width=0.48\textwidth]{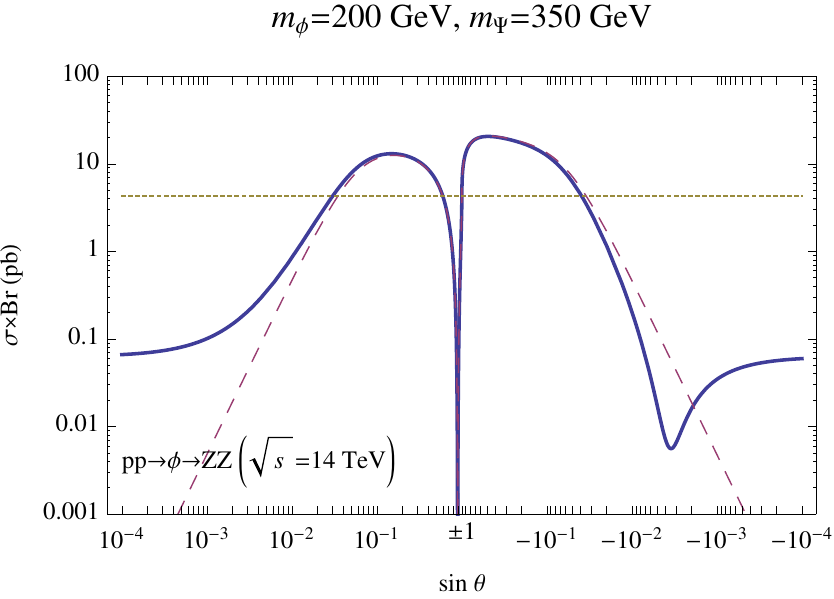} 
            \put(-108,-15){(a)}
             \put(115,-15){(b)}
     \includegraphics[width=0.48\textwidth]{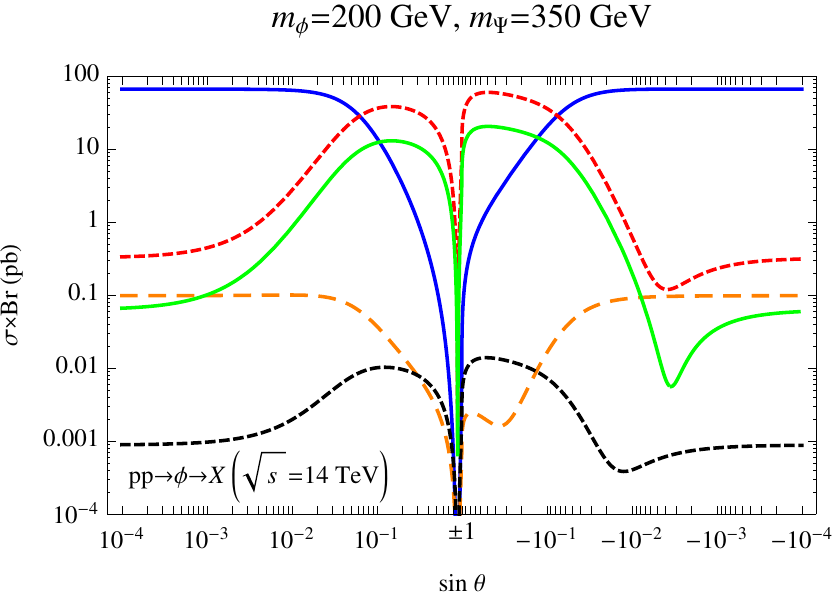}  
     \put(0,38){$\gamma Z$}
     \put(0,70){$ZZ$}
    \put(0,82){$\gamma \gamma$}
   \put(0,90){$WW$}
     \put(0,135){$gg$}
   \caption{(a) Cross section times branching ratio for $pp \rightarrow \phi \rightarrow ZZ$ at the LHC with $\sqrt{s} = 14$ TeV, taking  $m_\phi = 200$ GeV and $m_{10} = 350$ GeV.  The horizontal dotted line shows the result for a 200-GeV SM Higgs, and the dashed line gives the result when the electroweak charges of the loop particles are turned off.  (b)  Cross section times branching ratio for various final states.   }
   \label{fig:flakey}
\end{figure}

Just as having charged matter in the loop also  allows $\phi$ to decay into $\gamma \gamma$ even in the absence of mixing, it also allows decays into $\gamma Z$, $ZZ$ and $WW$ in that limit.  In Figure \ref{fig:flakey}a we see that the rate for $pp \rightarrow \phi \rightarrow ZZ$ with our benchmark matter content in the loop and a 200 GeV $\phi$ is essentially the same as if the matter were colored but not charged, unless the mixing is small.  As mentioned earlier, this is because the the decay through mixing is tree-level and tends to dominate.    The rate asymptotes to a non-zero value as the mixing is turned off, but for the parameters we have chosen, at least,  it is well below the rate for a SM Higgs of the same mass.  Sticking with the same matter content in the loop, in Figure \ref{fig:flakey}b we show $\sigma\times Br$ for $\gamma \gamma$, $\gamma Z$, $ZZ$ and $WW$ at the LHC with $\sqrt{s} = 14$ TeV.


\section{Higgs counterfeits, Higgs friends and effective $Z'$s}
\label{sec:effectivezprime}
What are the ingredients for either a friend or counterfeit scenario? They are quite simple: a new scalar, some heavy colored matter and some heavy charged matter. These are precisely the ingredients already present in the recently considered ``effective $Z'$'' scenario \cite{Fox:2011qd}. Here, rather than having SM fields charged directly under a $Z'$, the couplings arise from an effective operator
\be
({M^{-2})^{i}_j}\bar q_i \gamma_\mu q^j S^* D^\mu S \supset g' ({M^{-2})^{i}_j}\bar q_i \gamma_\mu q^j S^* Z^{\prime\mu} S~.
\label{eq:operator}
\ee
Such an operator can easily arise after integrating out heavy quarks (or leptons) charged under $SM\otimes G_{Z'}$, where $G_{Z'}$ is the (potentially non-Abelian group) under which the $Z'$ is the gauge field.

Taking a limiting case of a single flavor of heavy quarks, we have, as in \cite{Fox:2011qd},
\be
{\mathcal{L}}\supset -\mu Q Q^c - y S q Q^c + h.c.
\label{eq:hopterms}
\ee
The mixing arises when $S$ breaks the $U(1)$, $S = \langle S \rangle + s/\sqrt{2}$, giving mass eigenstates
\be
\tilde Q = \cos \theta Q + \sin \theta q \hskip 0.5in \tilde q = -\sin \theta Q + \cos \theta q,
\ee
where
\be
\sin \theta = \frac{y \vev{S}}{\sqrt{\mu^2+ y^2 \vev{S}^2}}
\label{eq:theta}
\ee
determines the mixing angle.  

If $s$ is produced through gluon-gluon fusion, we are immediately confronted by how it will decay. If $m_s < 2 M_{Z'}$, decays into $Z'$ will be kinematically inaccessible, and if  $m_s < M_{Z'}$ decays to SM particles through the $Z'$ will be four-body, and thus highly suppressed. If $s$ is lighter than the heavy quarks we are integrating out, the decay into SM states will be Yukawa suppressed, with a 
${\bar {\tilde q}}{\tilde q} s$ coupling equal to 
\cite{Fox:2011qd}
\be
g_{eff} \frac{m_q}{M_{Z'}}~,
\ee
where $m_q$ is the mass of the light quark and $g_{eff}$ is the effective coupling of the $Z'$ to matter. 
Note that the presence of this coupling depends on the heavy quarks being integrated out, so it may or may not be there for all quarks. A Higgs counterfeit can be realized by the inclusion of a $|S|^2|H|^2$ term, which will mix $s$ and $h$. If such a term is not present, the branching ratios into SM fermions can be quite small. In fact, if $m_s < 2 m_{\rm top}$, it is possible to boost the diboson decays in the presence of vectorlike heavy leptons, yielding a natural realization of a Higgs friend.  For this to work we need $m_s \lsim M_{Z'}$, to suppress $Z'$ decay modes.

Let us estimate the size of the $\gamma \gamma$ signal in the absence of  $s - h$ mixing.  The Yukawa coupling of $s$ to the heavy quarks can be worked out to be 
\begin{equation}
y = g_{eff} \frac{M_Q}{M_{Z'}},
\end{equation}
where $M_Q$ is the heavy quark mass.
The loop-induced $s \rightarrow \gamma \gamma$ signal is proportional to $ (y/M_Q)^2 = (g_{eff}/ M_{Z'})^2 $ for $M_Q \gg m_s/2$, which means that if all of the heavy particles are degenerate we can take
\begin{equation}
R_{friend} = R_{benchmark} \times K \times (g_{eff} \times 350 \;{\rm GeV}/ M_{Z'})^2,
\end{equation}
where $R_{benchmark}\simeq 2.5$ is the value of $R_{friend}$ in the absence of mixing for our benchmark point (with $s$ coupling to quarks only), and $K$ depends on charges and color factors.  

Taking $g_{eff} \sim 0.24$ and $M_{Z'} \sim 150$ GeV  to explain the $Wjj$ excess \cite{Fox:2011qd}, 
and setting $K = (5/3)^2$ because this model requires three copies because of $Q$ to complete it, we find $R_{friend} \sim 2$.  This will not be significantly suppressed by decays into $\bar b b$, because
decays into gluons are dominant for this parameter point. Taking $g_{eff} \sim 2$ as needed to produce the top $A_{FB}$ excess, the same $M_{Z'}$, and $K=(4/9)^2$ , the appropriate value given that this scenario requires a single Dirac $u^c$, we get $R_{friend} \sim 10$. 
Thus, for $Z'$s relevant for either the $Wjj$ excess or the top $A_{FB}$ excess, a Higgs friend would be a natural accompanying signal, with an appreciable size.

\section{Searches for new colored states} 
\label{sec:coloredstates}

Discovery of a Higgs counterfeit or friend may be preceded by discovery of the colored states that allow the production of the new resonance. While a thorough examination of the decay modes is beyond our scope, we present a summary of the existing limits and future strategies here.

Consider first the case in which $\Psi$ is a color-octet Majorana fermion.  The experimental lower bound on the mass of this ``gluino'' depends on how it decays.  If $\Psi$ is long lived ($\mu s-1000s$) ATLAS has a constraint of $\sim 560\,\gev$ \cite{Aad:2011yf}. 
Longer-lived octets run into cosmological constraints~\cite{Arvanitaki:2005fa}, from BBN~\cite{Kawasaki:2004qu}, the diffuse gamma ray background~\cite{Kribs:1996ac}, the CMB~\cite{Hu:1993gc,Feng:2003uy}, and searches for anomalously heavy isotopes~\cite{Smith:1982qu,Hemmick:1989ns}. 

A possible final state for a promptly decaying $\Psi$ is $q {\overline q} \chi$, where $\chi$ is a SM-singlet fermion.  If the SM neutrinos are of Dirac type $\chi$ could be an effectively massless right-handed neutrino. In this case Tevatron searches constrain the mass of $\Psi$ to be greater than $\sim 460$ GeV  \cite{Alwall:2008ve}.  Or, the $\chi$ particle might instead be a  massive fermion from the same hidden sector as $\phi$, in which case this constraint is weakened.  Earlier studies found that if the mass of $\chi$ is 200 GeV, for instance, all $\Psi$ masses above 205 GeV are consistent with Tevatron and early LHC results \cite{Alwall:2008ve, Alves:2010za}, however,  more recent LHC \cite{daCosta:2011qk, Chatrchyan:2011wb} results probably require larger masses.  In \cite{daCosta:2011qk} it is found that if $\chi$ particle is massless, $\Psi$ must be heavier than 500 GeV.   

Another possibility is that the octet decay is $\Psi \rightarrow q {\overline q} l$, which might be induced by the higher-dimension operator $\Psi D \overline{U} \overline{E}$. Leptoquark searches such as \cite{:2009gf} could be sensitive to this decay, but it is not obvious how published limits would be affected by the energy of $\Psi$ being shared among three particles rather than two, and by $\Psi$ being an  octet fermion as opposed to a color-triplet scalar or vector.  Bounds on leptoquarks moreover depend on the produced quark and lepton flavors.

In the case that the new matter can have renormalizable couplings such as $\Psi u^c H$  with SM matter (as is naturally the case in the friends scenario) and can decay promptly, the dominant constraints come from searches for fourth generations, putting a lower bound of $M_\psi \gtap 350\,\gev$ \cite{Lister:2008is,Aaltonen:2009nr,Flacco:2010rg}. We shouldn't expect vector-quark  constraints to be identical to those for a fourth-generation, and they will in any case be parameter dependent.  Furthermore, in the effective $Z'$ scenario, the heavy quarks can decay via off- or on-shell $Z'$s into three jet final states (or rather, a six-jet final state as they will be pair-produced). Absent model-dependent flavor tags, the limits on such scenarios are weak and we refer the reader to \cite{Fox:2011qd} for further discussion.

\section{Discussion}
\label{sec:discussion}
The discovery of the Higgs will be an important step towards whatever is next. In concert with the discovery of new states at the LHC or Tevatron, the mysteries of the weak scale will be slowly unveiled. But if we discover a Higgs, we must be certain that it is a Higgs. We have shown two different examples of scenarios where a ``Higgs'' could be discovered, either as a Higgs counterfeit or a Higgs friend. While in the latter case, a study of relative branching ratios might indicate that it is something quite different from a Higgs, in the former case, with the inaccessibility of the $H\rightarrow jj$ final state, even the branching ratios will naturally mimic those of the Higgs.

Such scenarios are simple, and easily realizable in physics beyond the SM.  Requiring the addition of only new colored matter, and a new scalar.  While a fundamental scalar invites the same radiative stability questions as the Higgs itself, we imagine that those are naturally solved by the same mechanism. Intriguingly, these scenarios arise naturally in the context of effective $Z'$ models, which necessitate new, vector-like matter, charged under the SM and some new group. The indications of new forces at the Tevatron, if confirmed, could well be accompanied by a Higgs counterfeit or Higgs friend at a comparable mass scale. With integrated luminosity increasing rapidly at the LHC, a discovery of any resonance would be a watershed moment, but  we must remember that only with a study of associated production modes, where the Higgs-$Z$ or Higgs-$W$ coupling is tested, will we truly know that we have found a Higgs.


\begin{acknowledgments}
The authors thank Kyle Cranmer, Bogdan Dobrescu, Roni Harnik, Graham Kribs, Joe Lykken and Adam Martin for many helpful discussions.   NW is supported by NSF grant PHY-0449818, as well as by the Ambrose Monell Foundation. DTS is is supported by NSF grant PHY-0856522. The authors thank Nima Arkani-Hamed for extensive discussions. The authors also thank the IAS and the Aspen Center for Physics for kind hospitality while parts of this work were undertaken. 
This work was inspired in part by the 2010 workshop "The Terascale at LHC 
0.5 and Tevatron" which was hosted by the University of Washington 
and supported by the University of Oregon DOE contract 
DE-FG02-96ER40969.
Fermilab is operated by Fermi Research Alliance, LLC, under Contract DE-AC02-07CH11359 with the United States Department of Energy.
\end{acknowledgments}


\appendix
\section{Appendix}
In this appendix we calculate the partial widths of $\phi$ and ${\tilde h}$ into $\gamma \gamma$, $\gamma Z$, $ZZ$, and $WW$.  We assume that $s =\cos \theta \phi - \sin\theta {\tilde h} $ has Yukawa couplings $y_i$ to Dirac fermions $\Psi_i$ with electric charges $Q_i$, isospins $T^3_i$, color  multiplicities $N^c_i$, and masses $m_i$.  
For $\phi \rightarrow \gamma \gamma$ we have \cite{Ellis:1975ap, Shifman:1979eb, Gavela:1981ri}
\begin{equation}
\begin{split}
\Gamma_{\phi}(\gamma \gamma) = 
\frac{G_F \alpha^2}{128\sqrt{2} \pi^3} m_{\phi}^3
&\left|
\cos\theta
\sum_i 
\left(
N^c_i Q_i^2 \frac{\sqrt{2} \;y_iv}{m_i}  A_{1/2}(\tau_i)
\right) \right. \\
& \quad\quad \quad\quad \left.-\sin \theta
\Biggl(
N_t^c Q_t^2 A_{1/2}(\tau_t) + A_1(\tau_W)
\Biggr)
\right|^2,
\end{split}
\end{equation}
where we take $v^2 = G_F^{-1}/(2\sqrt{2}) \simeq (174\; {\rm GeV})^2$ and define $\tau_i = m_\phi^2/(4 m_i^2)$, $\tau_t = m_\phi^2/(4 m_t^2)$,  and $\tau_W = m_\phi^2/(4 m_W^2)$.  The $A_{1/2}$ and $A_1$ functions are as defined in \cite{Djouadi:2005gi},
\begin{equation}
A_{1/2}(\tau) = 2 [\tau+(\tau-1) f(\tau)]\tau^{-2}
\end{equation}
\begin{equation}
A_{1}(\tau) = - [2 \tau^2 + 3 \tau+3(2\tau-1) f(\tau)] \tau^{-2}
\end{equation}
\begin{equation}
f(\tau)= 
\left\{
\begin{array}{ccc}
\arcsin^2 \sqrt{\tau} & & \tau \leq1 \\
 -\frac{1}{4}\left[ \log\frac{1+\sqrt{1-\tau^{-1}}}{1-\sqrt{1-\tau^{-1}}} - i\pi \right]^2 & & \tau>1
\end{array}
\right. .
\end{equation}
The partial width for $\phi \rightarrow \gamma Z$ is \cite{Cahn:1978nz,Bergstrom:1985hp}
\begin{equation}
\begin{split}
\Gamma_{\phi}(\gamma Z) &=
\frac{G_F^2 m_W^2 \alpha}{64 \pi^4} m_{\phi}^3 \left( 1-\frac{m_Z^2}{m_\phi^2}\right)^3
\left|
\cos\theta
\sum_i 
\left(
N^c_i Q_i \frac{\sqrt{2} \;y_iv}{m_i} \frac{4(T^3_i-Q_i s_w^2)}{c_w} A_{1/2}(\tau_i,\lambda_i)
\right) \right. \\
&\left.  -\sin \theta
\Biggl(
N_t^c Q_t \frac{2 T^3_t - 4 Q_t s_w^2}{c_w} A_{1/2}(\tau_t, \lambda_t) + A_1(\tau_W,\lambda_W)
\Biggr)\right|^2,
\end{split}
\end{equation}
where we define $\lambda_i = m_Z^2/(4 m_i^2)$,  $\lambda_t = m_Z^2/(4 m_t^2)$,  $\lambda_W = m_Z^2/(4 m_W^2)$, and 
\begin{equation}
A_{1/2}(\tau,\lambda)= I_1(\tau,\lambda) - I_2(\tau,\lambda)
\end{equation}
\begin{equation}
A_{1}(\tau,\lambda)= c_w \left\{
4\left(3-\frac{s_w^2}{c_w^2}\right)I_2(\tau,\lambda)+ \left[ 
\left(1+\frac{2}{\tau}\right)\frac{s_w^2}{c_w^2} - \left(5+\frac{2}{\tau} \right)
\right] I_1(\tau,\lambda)
\right\}
\end{equation}
\begin{equation}
I_1(\tau,\lambda)= \frac{1}{2(\lambda-\tau)}+\frac{1}{2(\lambda-\tau)^2}[f(\tau)-f(\lambda)]+\frac{\lambda}{(\lambda-\tau)^2}[g(\tau) - g(\lambda)]
\end{equation}
\begin{equation}
I_2(\tau,\lambda)= -\frac{1}{2(\lambda-\tau)}[f(\tau)-f(\lambda)].
\end{equation}
The function $f(\tau)$ is defined as above, and $g(\tau)$ is given as
\begin{equation}
g(\tau)= 
\left\{
\begin{array}{ccc}
\sqrt{\tau^{-1} - 1} \arcsin \sqrt{\tau} & & \tau \leq1 \\
\frac{\sqrt{1-\tau^{-1}}}{2}\left[ \log\frac{1+\sqrt{1-\tau^{-1}}}{1-\sqrt{1-\tau^{-1}}} - i\pi \right]^2 & & \tau>1
\end{array}
\right. .
\end{equation}
The partial width for $\phi \rightarrow ZZ$ can be expressed as
\begin{equation}
\Gamma_\phi(ZZ) = \frac{1}{32 \pi}\frac{1}{m_\phi}\sqrt{1-\frac{4 m_Z^2}{m_\phi^2}} 
\left(
2 \left| 
{\mathcal M}^{tree}_{++}+\sum_i {\mathcal M}^i_{++} 
\right|^2
+\left| 
{\mathcal M}^{tree}_{LL}+\sum_i {\mathcal M}^i_{LL} 
\right|^2
\right),
\end{equation}
where the tree-level amplitudes are
\begin{equation}
{\mathcal M}^{tree}_{++} =  \sin\theta \sqrt{2}  \frac{m_Z^2}{v}
\end{equation}
\begin{equation}
{\mathcal M}^{tree}_{LL} = -  \sin\theta \frac{1}{\sqrt{2}}  \left(\frac{m_\phi^2 - 2m_Z^2}{v} \right),
\end{equation}
and the contributions due to $\Psi_i$ loops are
\begin{equation}
{\mathcal M}^{i}_{++} = -\cos\theta \frac{2 \alpha }{\pi s_w^2 c_w^2} N^c_i (T^3_i-Q_i s_w^2)^2  m_i \;I_{++} (\tau_i,\lambda_i)
\label{eq:M++}
\end{equation}
\begin{equation}
{\mathcal M}^{i}_{LL} =\cos\theta \frac{2 \alpha }{\pi s_w^2 c_w^2} N^c_i (T^3_i-Q_i s_w^2)^2  m_i \;I_{LL} (\tau_i,\lambda_i).
\label{eq:MLL}
\end{equation}
We neglect the contributions from top and $W$ loops, which are tiny compared to the tree-level contributions and also proportional to $\sin\theta$.  The integrals appearing in the $\Psi_i$ contributions are
\begin{equation}
I_{++}(\tau,\lambda) = \int_0^1 dx \int_0^{1-x} dy  \left[ \frac{4x(1-2x) \lambda+4 y(1-2y )\lambda +4(2xy-1/2)(\tau-2 \lambda)}{1-4x(1-x) \lambda-4y(1-y)\lambda-4xy(\tau-2\lambda)-i\epsilon} \right]
\end{equation}
\begin{equation}
I_{LL}(\tau,\lambda)  =  \int_0^1 dx \int_0^{1-x}\! dy  \;
\left[\frac{- 4 \lambda(1-4xy)+2(\tau-2\lambda)[x(1-2x) +y(1-2y)]}
{1-4x(1-x) \lambda-4y(1-y)\lambda-4xy(\tau-2\lambda)-i\epsilon}\right]
\end{equation}
If any of the $\tau_i$  are greater than one, ${\mathcal M}^{i}_{++}$ and ${\mathcal M}^{i}_{LL} $ acquire imaginary parts.    

The partial width for $\phi \rightarrow W^+ W^-$ can be obtained from that for $\phi \rightarrow ZZ$ straightforwardly.  Neglecting mass splittings within $SU(2)$ multiplets in the $\Psi_i$,  the amplitude contributions from a Dirac fermion in a given $SU(2)$ representation $r$ is given by Equations~(\ref{eq:M++})   and (\ref{eq:MLL}) with the replacement $m_Z \rightarrow m_W$ and  $\alpha (T^3_i-Q_i s_w^2)^2/(s_w^2 c_w^2) \rightarrow C(r) \alpha/(s_w^2)$, with $C(r) = 1/2$ for doublets.  Elsewhere one needs to multiply the partial width by an overall factor of 2 and replace $m_Z \rightarrow m_W$ wherever it appears.  

The formulae for the partial widths of ${\tilde h}$ into $\gamma \gamma$, $\gamma Z$, $ZZ$ and $W^+ W^-$ are given by the $\phi$ partial widths given above, with the replacements $\cos \theta \rightarrow \sin\theta$, $\sin\theta \rightarrow -\cos\theta$, and $m_\phi \rightarrow m_{\tilde h}$.

\bibliography{imposter}
\end{document}